\documentclass[floatfix,aps,prx,twocolumn,superscriptaddress]{revtex4-2}

\usepackage[T1]{fontenc}
\usepackage{lipsum}  
\usepackage{graphicx}
\usepackage[english]{babel}
\usepackage{upgreek}
\usepackage{amsmath}
\usepackage{tabls}
\usepackage[colorlinks=true,citecolor=blue,linkcolor=red,urlcolor=blue]{hyperref}
\usepackage[normalem]{ulem}
\usepackage{xcolor}
\usepackage{float}

\makeatletter
\def\bbl@set@language#1{%
	\edef\languagename{%
		\ifnum\escapechar=\expandafter`\string#1\@empty
		\else\string#1\@empty\fi}%
	\@ifundefined{babel@language@alias@\languagename}{}{%
		\edef\languagename{\@nameuse{babel@language@alias@\languagename}}%
	}%
	\select@language{\languagename}%
	\expandafter\ifx\csname date\languagename\endcsname\relax\else
	\if@filesw
	\protected@write\@auxout{}{\string\select@language{\languagename}}%
	\bbl@for\bbl@tempa\BabelContentsFiles{%
		\addtocontents{\bbl@tempa}{\xstring\select@language{\languagename}}}%
	\bbl@usehooks{write}{}%
	\fi
	\fi}
\newcommand{\DeclareLanguageAlias}[2]{%
	\global\@namedef{babel@language@alias@#1}{#2}%
}
\makeatother

\newcommand\varpm{\mathbin{\vcenter{\hbox{%
  \oalign{\hfil$\scriptstyle+$\hfil\cr
          \noalign{\kern-.3ex}
          $\scriptscriptstyle({-})$\cr}%
}}}}
\newcommand\varmp{\mathbin{\vcenter{\hbox{%
  \oalign{$\scriptstyle({+})$\cr
          \noalign{\kern-.3ex}
          \hfil$\scriptscriptstyle-$\hfil\cr}%
}}}}

\DeclareLanguageAlias{en}{english}

\begin{document}

\title{Nuclear  spin readout in a cavity-coupled hybrid quantum dot-donor system}

\author{Jonas Mielke}
\affiliation{Department of Physics, University of Konstanz, D-78457 Konstanz, Germany}
\author{Jason R. Petta}
\affiliation{Department of Physics, Princeton University, Princeton, New Jersey 08544, USA}
\author{Guido Burkard}
\affiliation{Department of Physics, University of Konstanz, D-78457 Konstanz, Germany}


\begin{abstract}
Nuclear spins show long coherence times and are well isolated from the environment, which are properties making them promising for quantum information applications. 
Here, we present a method for nuclear spin readout by probing the transmission of a microwave resonator. We consider a single electron in a silicon quantum dot-donor device interacting with a microwave resonator via the electric dipole coupling and subjected to a homogeneous magnetic field and a transverse magnetic field gradient. In our scenario, the electron spin interacts with a $^{31}\mathrm{P}$ defect nuclear spin via the hyperfine interaction. 
We theoretically investigate the influence of the P nuclear spin state on the microwave transmission through the cavity and show that nuclear spin readout is feasible with current state-of-the-art devices. Moreover, we identify optimal readout points with strong signal contrast to facilitate the experimental implementation of nuclear spin readout. 
Furthermore, we investigate the potential for achieving coherent excitation exchange between a nuclear spin qubit and cavity photons.
\end{abstract}


\maketitle

\section{Introduction}
Nuclear spins are promising candidates for quantum information applications due to their long coherence times \cite{muhonen2014,pla2013} that can even be observed up to room temperature \cite{saeedi2013}.
However, the small gyromagnetic ratio of nuclear spins that renders them well isolated from the environment and thus underlies their robust coherence also leads to long gate operation times compared to those reported for electron spins. Therefore, nuclear spins will most likely find their application as quantum memories \cite{maurer2012,steger2012}, either for pure storage or as buffers for quantum computation purposes. In many cases, it is beneficial to perform readout directly on the nuclear spin instead of coherently transferring the information to another system, e.g. an electron spin \cite{freer2017}, before this system is read out. Particularly with regard to scalable quantum computing devices, readout relying on electrical means is favourable over methods depending on ac-magnetic fields.  

To a large extent driven by the microelectronics industry, the manufacturing of nanoscale semiconductor devices has matured during the last decades, and silicon-based devices have a particularly high potential for scaling. Moreover, isotopically purified $^{28}\mathrm{Si}$ material containing predominantly nuclear spin $0$ atoms can be produced and thus provides an excellent host material for spin qubits based on the electron spin or single impurity nuclear spins \cite{loss1998,kane1998,tyryshkin2012,zwanenburg2013,awschalom2013,veldhorst2014,morton2008,freer2017}.

Cavity quantum electrodynamics (cQED) has been successfully used for charge-photon  \cite{mi2017,stockklauser2017,bruhat2018} and spin-photon coupling \cite{kubo2010,laflamme2012,tosi2014,mi2018,benito2017,landig2018,samkharadze2018,cubaynes2019}, as well as the detection of photons \cite{schuster2007}. 
Moreover, cQED and gate reflectometry lend themselves for qubit readout \cite{blais2004,didier2015,didier2015a,govia2017,mi2017,danjou2019,zheng2019,west2019,crippa2019}. Nevertheless, it is an open question whether or not single nuclear spins could be detected via a cavity or coupled to cavity photons.
Among the before mentioned achievements in cQED it is particularly noteworthy that the strong coupling regime is accessible for the spin of a single electron in a $\mathrm{Si}$ double quantum dot (DQD) subject to a magnetic field gradient where spin photon coupling emerges due to
\begin{figure}[H]
	\includegraphics[width=\columnwidth]{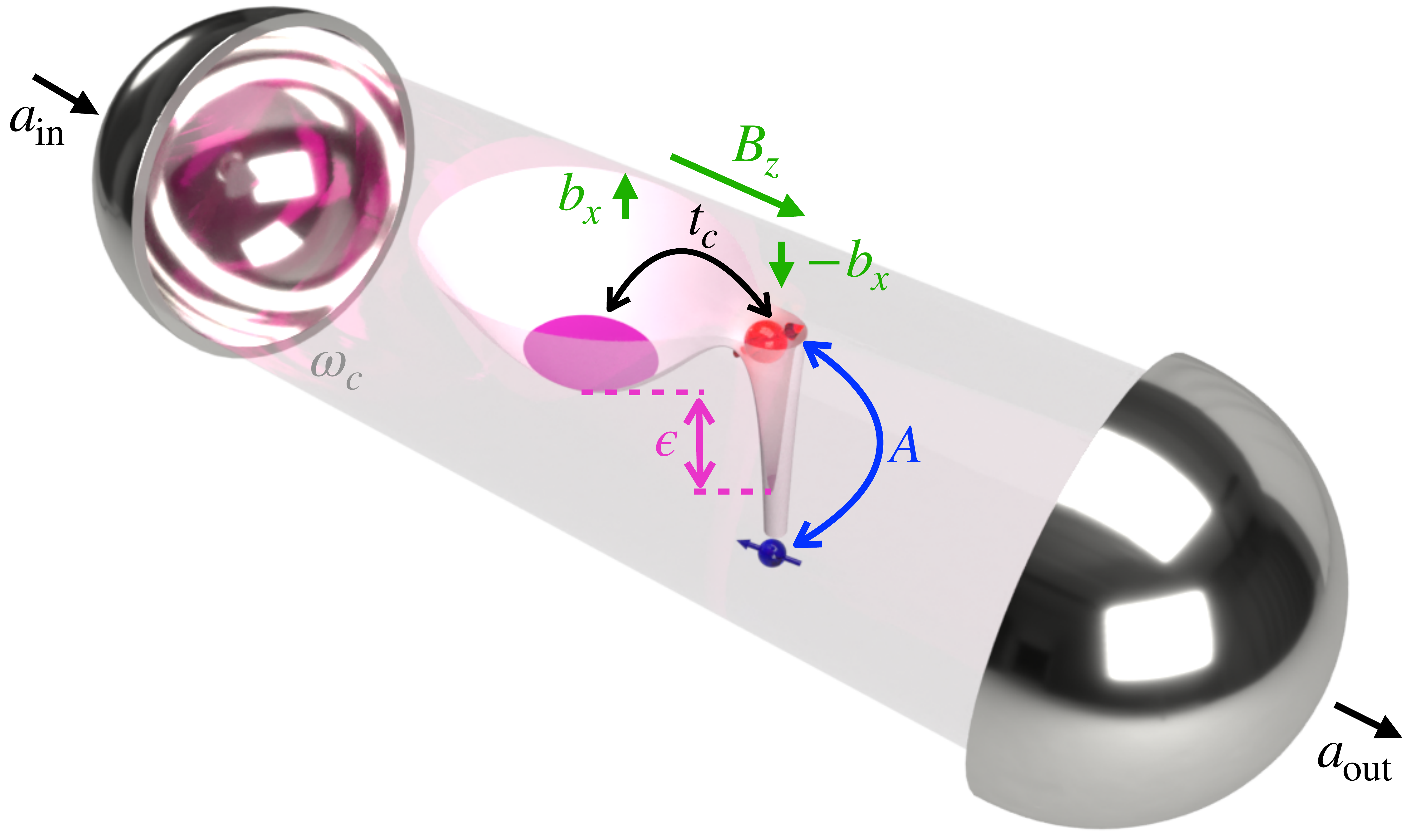}
	\caption{ 
	QD-donor system coupled to a single-mode microwave cavity.  
	The cavity transmission $A_c=\langle a_{\mathrm{out}}\rangle/\langle a_{\mathrm{in}}\rangle$ reveals the nuclear spin state. The QD-donor energy levels are detuned by an amount $\epsilon$ and hybridized by tunnel coupling $t_c$. The spin of the confined electron is subject to a homogeneous magnetic field $B_z$ and a gradient field $b_x$ perpendicular to $B_z$, while the electron charge is coupled to a single mode ($\omega_c$) of a microwave cavity with electric-dipole coupling strength $g_c$. The electron interacts with the nuclear spin of the implanted donor via the hyperfine interaction $A$.}
	\label{fig:model}
\end{figure}
\noindent an effective spin orbit coupling caused by the the combination of the electric dipole interaction and spin-charge hybridization \cite{mi2018,benito2017}. The same mechanism can also be used to realize a flopping-mode spin qubit with full electrical spin control via electric dipole spin resonance  (EDSR) \cite{benito2019,croot2020}. In this system, a longitudinal magnetic field gradient leads to a shift of the phase and amplitude response of the cavity transmission depending on the strength of the field gradient \cite{croot2020}. 
Motivated by this observation we consider a lateral architecture consisting of a quantum dot (QD) in a planar Si/SiGe structure and a single $^{31}\mathrm{P}$ donor implanted in the Si host material. While this system has been successfully operated in the multi electron regime \cite{harvey-collard2017}, we consider the single electron regime to form a flopping-mode electron spin qubit (Fig.~\ref{fig:model}).

As a consequence, if the electron is confined to the donor, it
couples to the donor nuclear spin via the hyperfine interaction. In this configuration, we expect the donor to generate a nuclear spin state dependent Overhauser field. This field constitutes a longitudinal magnetic field gradient that leads to a nuclear spin state and detuning dependent shift of the electron spin transition frequency. Therefore, the cavity response essentially probing the EDSR frequency is expected to shift accordingly.
Our detailed discussion of the expected characteristics in the cavity transmission indicates that the observable signature of the strong electron spin photon coupling \cite{benito2017,mi2018} is indeed significantly altered by the state of the nuclear spin and could therefore be used for nuclear spin state readout. This prediction is verified by calculating the cavity transmission using input-output theory.

Moreover, we investigate the effective excitation conserving nuclear spin photon coupling and find that our suggested method for nuclear spin readout does not require strong nuclear spin photon coupling. 

This article is organized as follows:  The following Sec.~\ref{sec:model} contains a discussion of the model of the QD-donor system coupled to a cavity mode. In Sec.~\ref{sec:nuclearspinreadout} we predict the impact of the donor nuclear spin state on the cavity transmission, and verify our expectation by calculating the cavity transmission using input-output theory. Section~\ref{sec:nuclear_spin-photon_coupling} contains the derivation of an effective Hamiltonian describing the nuclear spin dynamics followed by a discussion of the emerging effective nuclear spin photon coupling. Finally, we summarize our results in Sec.~\ref{sec:conclusion}.

\section{Theoretical Model \label{sec:model}}
We consider a lateral QD-donor system fabricated in isotopically enriched $^{28}$Si. QD and donor are aligned along the $z$-axis, such that a single electron can either be localized in the QD on the left or the  donor on the right by adjusting the  QD-donor energy level detuning $\epsilon$. QD-donor tunnel coupling $t_c$ results in charge hybridization near $\epsilon = 0$. The proposed experimental setup, including the various interactions, is sketched in Fig.~\ref{fig:model}. 
The detuning $\epsilon$ determined by the energy difference between QD and donor can be controlled by applying an electric field in $z$-direction and by tuning the gates defining the QD confinement potential.
In the presence of a homogeneous magnetic field $B_z$ and a magnetic field gradient $b_x$ perpendicular to $B_z$, the
QD-donor system in the single electron configuration can be modelled by the Hamiltonian
\begin{eqnarray}
\widetilde{H}_0=\tfrac{1}{2}\left(\epsilon \widetilde{\tau}_z +2 t_c \widetilde{\tau}_x+B_z \sigma_z + b_x \sigma_x \widetilde{\tau}_z\right),
\label{eq:H0tilde}
\end{eqnarray}
with $\widetilde{\tau}_i$ and $\sigma_i$ the Pauli operators in position and electron spin space, respectively. The interaction between the nuclear spin and the magnetic field is neglected because it is roughly three orders of magnitude smaller than all other relevant energy scales \cite{tosi2017}. The magnetic fields $B_z$ and $b_x$ are given in units of energy and energy units are chosen such that $\hbar=1$. The donor ground and first excited state are energetically separated by $\gtrsim 2.5\,\mathrm{meV}$ taking into account strain effects due to the Si/SiGe interface \cite{usman2015}. On the other hand, a low-lying excited state, the excited valley state, is present in Si QDs. However, valley splittings $\gtrsim50\,\upmu\mathrm{eV}$ observed in recent devices \cite{mi2017b,borjans2021} together with the possibility to operate the QD-donor system at temperatures $\lesssim 50\,\mathrm{mK}$ allow for a negligible population of the excited valley state. Hence, the valley degree of freedom can be neglected in our model.

Electric dipole interactions allow to couple the electron in the
QD-donor system to microwave resonator photons, described by the coupling Hamiltonian
\begin{eqnarray}
\widetilde{H}_I=g_c\widetilde{\tau}_z(a+a^{\dagger}),
\end{eqnarray}
where $a$ and $a^{\dagger}$ are the bosonic cavity photon annihilation and creation operators of the relevant cavity mode, respectively. 
The charge-photon coupling strength for a DQD has been found to be on the order of $g_c/2\pi \approx 30$ to $40\,\mathrm{MHz}$ \cite{mi2018,petersson2012a}. In the QD-donor scenario we expect it to be $\approx 1/3$ of the DQD value, as discussed in Appendix~\ref{appendix:QD-donor system}.
The Hamiltonian for the cavity mode with frequency $\omega_c$ is given by $\widetilde{H}_{\mathrm{cav}}=\omega_c a^{\dagger}a$.
If the electron is confined to the donor, electron spin and $^{31}\mathrm{P}$ donor nuclear spin couple via the hyperfine interaction. The hyperfine interaction strength $A=117 \,\mathrm{MHz}$ \cite{steger2011,feher1959} present in bulk Si is significantly reduced to $A\approx 25\,\mathrm{MHz}$ in the Si quantum well of a Si/Si$_{0.7}$Ge$_{0.3}$ heterostructure due to strain effects caused by the Si- and SiGe-lattice mismatch \cite{huebl2006,usman2015}. On the other hand, the donor is ionized and the electron does not interact with the donor nuclear spin if it occupies the left QD. Therefore, we can represent the electron spin nuclear spin interaction as
\begin{eqnarray}
\widetilde{H}_{e-n}=\frac{A}{8}\vec{\sigma}\cdot \vec{\nu}\cdot (1-\widetilde{\tau}_z),
\label{eq:He-n}
\end{eqnarray}
with $\vec{\nu}=(\nu_x,\nu_y,\nu_z)^T$ and $\nu_i$ the nuclear spin Pauli operators. The factor $(1-\widetilde{\tau}_z)/2$ is a projection on the subspace with the electron bound to the donor.

\begin{figure}[ht]
	\centering
	\includegraphics[width=\columnwidth]{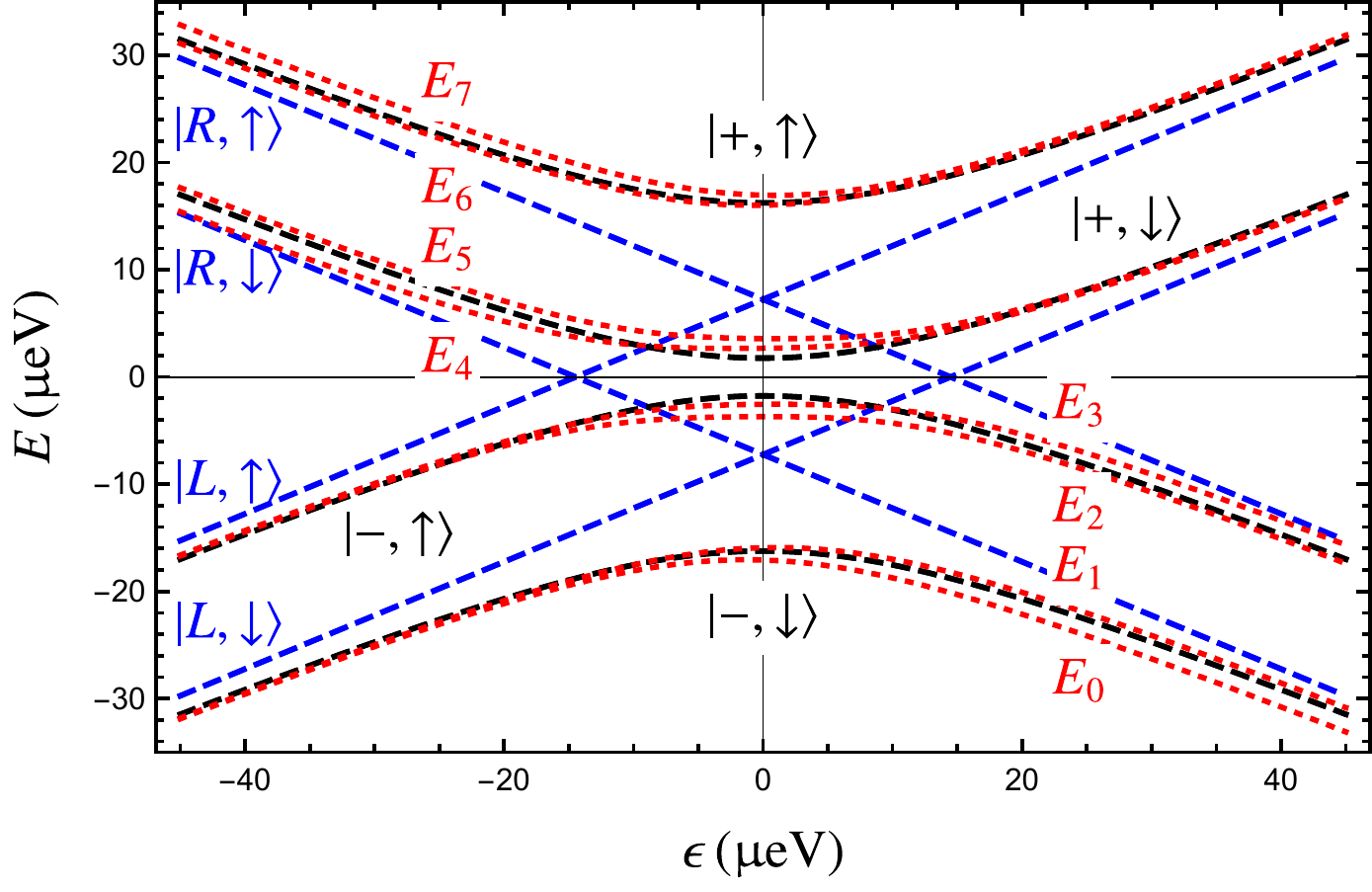}
	\caption{Energy spectrum of the QD-donor Hamiltonian $\widetilde{H}_{0}$ as a function of the
	QD-donor detuning parameter $\epsilon$ for $B_z=14.5\,\upmu\mathrm{eV}$. The dashed blue lines show the energy levels for $t_c=b_x=0$ with the corresponding eigenstates denoted by the electronic position in the
	QD-donor system, $L$ and $R$, and its spin projection $(\downarrow/\uparrow)$ along the $z-$axis. Switching on the tunnel coupling $t_c$ (here $t_c=9\,\upmu\mathrm{eV}$) mixes the $L$ and $R$ states with the same spin resulting in the bonding $(-)$ and antibonding $(+)$ molecular orbital states, whose energies are given by the dashed black lines. Finally, the dotted red energy levels include the hyperfine coupling to the nuclear spin (eigenenergies $E_0,\ldots,E_7$ of $\widetilde{H}_{\rm sys}$). For this plot, we have used exaggerated values $b_x=5\,\upmu\mathrm{eV}$ and $A=1\,\mathrm{GHz}$ to emphasize the effects of the magnetic field gradient and the hyperfine interaction.}
	\label{fig:energylevels}
\end{figure}

Signatures of the electron spin-photon coupling can be observed in the cavity transmission \cite{mi2018,benito2017}. We now investigate whether these signatures will be altered in the presence of a nuclear spin interacting with the electron spin via the hyperfine interaction and whether the combined spin-photon and hyperfine interactions have a potential application for nuclear spin readout. 

To calculate the cavity response, we first transform the total Hamiltonian to the eigenbasis $|\pm\rangle$ 
of $\epsilon \widetilde{\tau}_z/2 +t_c \widetilde{\tau}_x$,
with the electron position expressed in terms of antibonding ($+$) and bonding ($-$) molecular orbital states because these basis states are a good approximation for the eigenstates $|n\rangle$ with corresponding energies $E_n$ of $\widetilde{H}_{\rm sys}=\widetilde{H}_0+\widetilde{H}_{\mathrm{e-n}}$ as illustrated in Fig.~\ref{fig:energylevels}. Then, the Hamiltonian $\widetilde{H}=\widetilde{H}_{\rm sys}+\widetilde{H}_{I}+\widetilde{H}_{\mathrm{cav}}$ can be written as the sum of a diagonal part $H_0$ and an off-diagonal perturbation $V$ as $H=H_0+V$, with 
\begin{align}
H_0=&\tfrac{1}{2}\left(\Omega \tau_z + B_z \sigma_z  \right)+ 
\tfrac{1}{8} A \sigma_z \nu_z \left(1-\sin\theta\, \tau_z\right) + \omega_c a^{\dagger} a,
\label{eq:H0}\\
V=&\frac{b_x}{2}\left(\sin\theta\,\tau_z -\cos\theta\, \tau_x \right)\sigma_x + \frac{A}{8}\cos\theta(\sigma_z \nu_z) \,\tau_x \nonumber\\
&+\frac{A}{8}(\sigma_x \nu_x + \sigma_y \nu_y) \left(1- 
\sin\theta\,\tau_z + \cos\theta\,\tau_x\right)\nonumber\\
&+ g_c (\sin\theta\,\tau_z-\cos\theta\,\tau_x)(a + a^{\dagger}),
\label{eq:V}
\end{align}
where $\tau_i$ are Pauli operators acting on the space of bonding ($+$) and antibonding ($-$) orbitals, i.e. $\tau_z |\pm\rangle=\pm |\pm\rangle$.
Moreover, we introduce the orbital energy $\Omega=\sqrt{\epsilon^2+
	4t_c^2}$ and the orbital mixing angle $\theta=\arctan(\epsilon /2 t_c)$. 
$H_0$ is diagonal with respect to the basis $\{|\pm,\downarrow (\uparrow), \Downarrow (\Uparrow),n\rangle\}$ indicating the orbital state of the electron $(\pm)$, the electron spin state ($\downarrow,\uparrow$), the nuclear spin state ($\Downarrow,\Uparrow$) and the number of photons in the cavity mode $(n)$, respectively, while $V$ is purely off-diagonal in this basis. 
In order to predict the impact of the nuclear spin on the cavity transmission, we derive an effective Hamiltonian for the lower orbital subspace defined by the projection operator 
$P_0=(1-\tau_z)/2$,
that projects on the subspace spanned by the states 
${|-,\downarrow,\Uparrow,n\rangle},{|-,\downarrow,\Downarrow,n\rangle},{|-,\uparrow,\Uparrow,n\rangle},{|-,\uparrow,\Downarrow,n\rangle}$ with $n=0,1,2,\ldots $.
 
 As a next step, we apply a Schrieffer-Wolff transformation to decouple the subspaces defined by the projection operators $P_0$ and $Q_0=1-P_0$ \cite{bravyi2011}, 
to find the effective Hamiltonian $H_{\mathrm{eff}}=e^SHe^{-S}$,
and follow the perturbative method presented in \cite{bravyi2011} to determine the block off-diagonal and antihermitian generator $S$ defining the unitary transformation $e^S$.
If one chooses the ansatz $S=\sum_{n=1}^{\infty}S_n$ with $S_n\sim V^n$, the first contribution $(S_1)$ must obey the relation \cite{bravyi2011}
$[H_0,S_1]= P_0VQ_0+Q_0VP_0$.
This relation together with the commutation relations of the Pauli operators and the bosonic photon operators allows us to determine $S_1$.
The knowledge of $S_1$ is in turn sufficient to compute the effective Hamiltonian for the subspace defined by $P_0$ up to second order in the perturbation $V$ \cite{bravyi2011},
\begin{align}
H_{e}=&P_0H_0P_0+P_0VP_0\nonumber\\
&+\tfrac{1}{2}P_0[S_1,P_0VQ_0+Q_0VP_0]P_0.
\label{eq:Heffeformula}
\end{align}
The explicit form of $H_{e}$ is presented in Appendix \ref{appendix:effective_Hamiltonian}. However, for the following discussion it is essential to determine transition frequencies as precisely as possible. To this end, we transform $H_{e}$ to a basis accounting for the electron spin mixing due to the magnetic field gradient with the basis states
\begin{align}
|\tilde{\downarrow},\Uparrow(\Downarrow)\rangle=&\cos \frac{\phi}{2}|-,\downarrow,\Uparrow(\Downarrow)\rangle
-\sin\frac{\phi}{2}|-,\uparrow,\Uparrow(\Downarrow)\rangle,\\
|\tilde{\uparrow},\Uparrow(\Downarrow)\rangle=&\sin\frac{\phi}{2}|-,\downarrow,\Uparrow(\Downarrow)\rangle
+\cos\frac{\phi}{2}|-,\uparrow,\Uparrow(\Downarrow)\rangle,
\end{align}
defined by the electron spin mixing angle $\phi$ via
\begin{align}
\tan\phi=\frac{-b_x \sin \theta}{B_z\left( 1-\cos^2\theta\left(\tfrac{A^2}{8}+b_x^2\right)\tfrac{1}{\Omega^{2}-B_z^{2}}\right)}.
\end{align}
Since, here, $b_x\ll B_z$ the electron spin mixing angle is small and therefore the states $|\tilde{\downarrow}(\tilde{\uparrow})\rangle$ are predominantly the electron spin states $|\downarrow(\uparrow)\rangle$ up to small contributions of the opposite electron spin state. Hence, in the following we refer to $|\tilde{\downarrow}(\tilde{\uparrow})\rangle$ as the electron spin states. 
The diagonal part of the transformed Hamiltonian reads
\begin{align}
H_{e,0}=& \tfrac{1}{2}\left(E_{\tilde{\sigma}}+\delta E_{\tilde{\sigma}}\nu_z\right)\tilde{\sigma}_z+ \tfrac{1}{2} E_{\nu}\nu_z+\widetilde{\omega}_c a^\dagger a,
\label{eq:Heffdiagonalmain}
\end{align}
with the Pauli operators $\widetilde{\sigma}_i$ operating on the $|\tilde{\downarrow}(\tilde{\uparrow})\rangle$ states and 
\begin{align}
    &E_{\tilde{\sigma}}=\sqrt{{B_z^2}\left(1-\frac{\left(A^2+8b_x^2\right)\cos^2\theta}{8(\Omega^{2}-B_z^{2})}\right)^2+\left(b_x\sin\theta\right)^2},\\
    &\delta E_{\tilde{\sigma}}=\left(\frac{A}{4}(1\!+\!\sin\theta)
    +\frac{\Omega}{B_z}E_\nu\right)\cos\phi
    \nonumber\\
    &\quad\quad\quad -\frac{Ab_x\cos\theta B_z}{8(\Omega^{2}-B_z^{2})}\sin\phi,\\
    &E_{\nu}=\frac{A^2\cos^2\theta\,{B_z}}{{16(\Omega^{2}}-B_z^{2})},\\
    &\widetilde{\omega}_c=\omega_c-2\frac{g_c^2\cos^2\theta\,B_z}{\Omega^{2}-\omega_c^{2}},
\end{align}
as derived in Appendix \ref{appendix:effective_Hamiltonian}. 
Since the signatures of the electron spin-photon coupling that we expect to change  due to the nuclear spin are observed close to resonance between the electron spin transition and the resonator \cite{benito2017,mi2018}, it is justified to assume $E_{\tilde{\sigma}}\approx\widetilde{\omega}_c$. Under this assumption we can apply the rotating wave approximation (RWA) retaining terms rotating with frequencies $\ll E_{\tilde{\sigma}}\approx\widetilde{\omega}_c$ and find that the nondiagonal part of the transformed Hamiltonian comprises interactions between the electron spin and the nuclear spin of the  QD-donor system and the cavity mode
\begin{align}
H_{e,\mathrm{cav-int}}=
&- g_{\tilde{\sigma}\nu}\sin^2\frac{\phi}{2}\left(\tilde{\sigma}^+ \nu^+ a +\tilde{\sigma}^- \nu^- a^{\dagger}\right)\nonumber\\
&+g_{\tilde{\sigma}\nu}\cos^2\frac{\phi}{2}\left(\tilde{\sigma}^+ \nu^- a +\tilde{\sigma}^- \nu^+ a^{\dagger}\right)\label{eq:HeffeintRWA}\\
&+\left(g_{\tilde{\sigma}}\cos\phi+\delta g_{\tilde{\sigma}}\sin\phi\,\nu_z\right)\left(\tilde{\sigma}^+a+\tilde{\sigma}^- a^{\dagger}\right),\nonumber
\end{align}
with the explicit forms of the spin-photon couplings $g_{\tilde{\sigma}\nu}, g_{\tilde{\sigma}}, \delta g_{\tilde{\sigma}}$
given in Appendix \ref{appendix:effective_Hamiltonian}.
The interaction terms in \eqref{eq:HeffeintRWA} are of particular interest since one can expect to see signatures of these interactions in the transmission. However, the terms in the first line are negligible for $\phi\ll 1$. 

The terms in the second line of \eqref{eq:HeffeintRWA} incorporate a flip of both the nuclear spin and the electron spin if the two are antialigned  with the concomitant creation or annihilation of a cavity photon. This coupling emerges due to the combined effect of the dipole operator coupling the states $|-,\uparrow,\Downarrow(\Uparrow)\rangle$ and $|+,\uparrow,\Downarrow(\Uparrow)\rangle$, and the hyperfine interaction between the states $|+,\uparrow,\Downarrow\rangle$ and $|-,\downarrow,\Uparrow\rangle$. Thus, the interaction persists in the absence of the magnetic field gradient and has already been observed and analyzed in setups without such a gradient. The interaction can be used to control the flip-flop qubit and to construct gates between two such qubits \cite{tosi2017}, while the combination with an oscillating magnetic field allows for controlling the nuclear spin qubit and implementing a nuclear spin two-qubit gate \cite{tosi2018}. 

\begin{figure}[ht]
	\centering
	\includegraphics[width=\columnwidth]{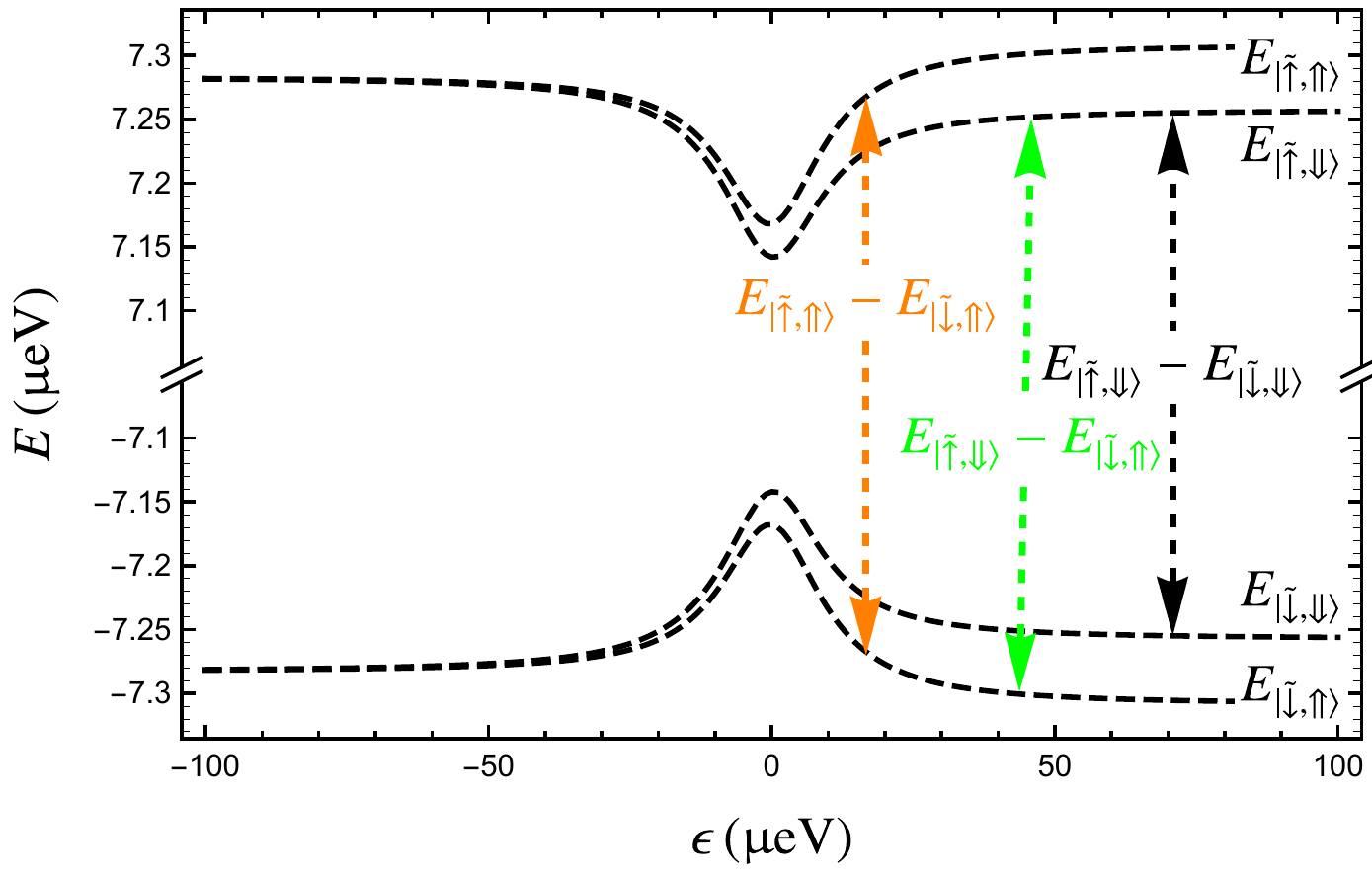}
	\caption{Hyperfine-split electronic energy levels (\ref{eq:energyexpectationvaluesdown}) and (\ref{eq:energyexpectationvaluesup}) corresponding to the four states $|\tilde{\downarrow},\Uparrow\rangle$, $|\tilde{\downarrow},\Downarrow\rangle$, $|\tilde{\uparrow},\Uparrow\rangle$ and $|\tilde{\uparrow},\Downarrow\rangle$ as a function of the QD-donor detuning parameter $\epsilon$. The highlighted transition frequencies determine the location of the observable signatures in Fig.~\ref{fig:cavityTransmission}. The  parameters used for this plot are $B_z=\omega_c/2\pi=3.5\,\mathrm{GHz}$, $g_c/2\pi=13\,\mathrm{MHz}$, $t_c=9\,\upmu\mathrm{eV}$, $b_x=1.62\,\upmu\mathrm{eV}$, and $A=25\,\mathrm{MHz}$.}
	\label{fig:energyspectrumheffdiag}
\end{figure}

On the other hand, the combined effect of the magnetic field gradient, giving rise to the coupling between the states $|+,\uparrow,\Downarrow(\Uparrow)\rangle$ and $|-,\downarrow,\Downarrow(\Uparrow)\rangle$, and the dipole operator leads to the terms in the third line that describe a flip of the electron spin  accompanied by the annihilation or creation of a cavity photon, while the state of the nuclear spin remains unchanged. These two different types of interaction cause a hybridization of the QD-donor system and the cavity mode when the transition in the QD-donor system is close to resonance with the cavity mode. Since the resulting hybrid states have a significant impact on the cavity transmission we inspect the energy expectation values of the QD-donor system states involved in the respective transitions. 
The energy expectation values of the four basis states defining the lower orbital subspace can be easily read off from \eqref{eq:Heffdiagonalmain}:
\begin{align}
&E_{|\tilde{\downarrow},\Uparrow(\Downarrow)\rangle}=-\frac{E_{\tilde{\sigma}}}{2}\varmp\frac{\delta E_{\tilde{\sigma}}}{2}\varpm\frac{E_{\nu}}{2},\label{eq:energyexpectationvaluesdown}\\
&E_{|\tilde{\uparrow},\Uparrow(\Downarrow)\rangle}= \frac{E_{\tilde{\sigma}}}{2}\varpm\frac{\delta E_{\tilde{\sigma}}}{2}\varpm\frac{E_{\nu}}{2}, 
\label{eq:energyexpectationvaluesup}
\end{align}
and we immediately find the transition frequencies for electron spin flips with a fixed nuclear spin state
\begin{align}
E_{|\tilde{\uparrow},\Uparrow(\Downarrow)\rangle}-E_{|\tilde{\downarrow},\Uparrow(\Downarrow)\rangle}=E_{\tilde{\sigma}}\varpm\delta E_{\tilde{\sigma}},
\label{eq:electrontransitionfrequency}
\end{align}
as well as the transition frequency for the electron spin-nuclear spin flip-flop
\begin{align}
E_{|\tilde{\uparrow},\Downarrow\rangle}-E_{|\tilde{\downarrow},\Uparrow\rangle}=E_{\tilde{\sigma}}-E_{\nu}.
\label{eq:ekectronnuclearflipflop}
\end{align}
\begin{figure*}[ht]
	\includegraphics[width=\textwidth]{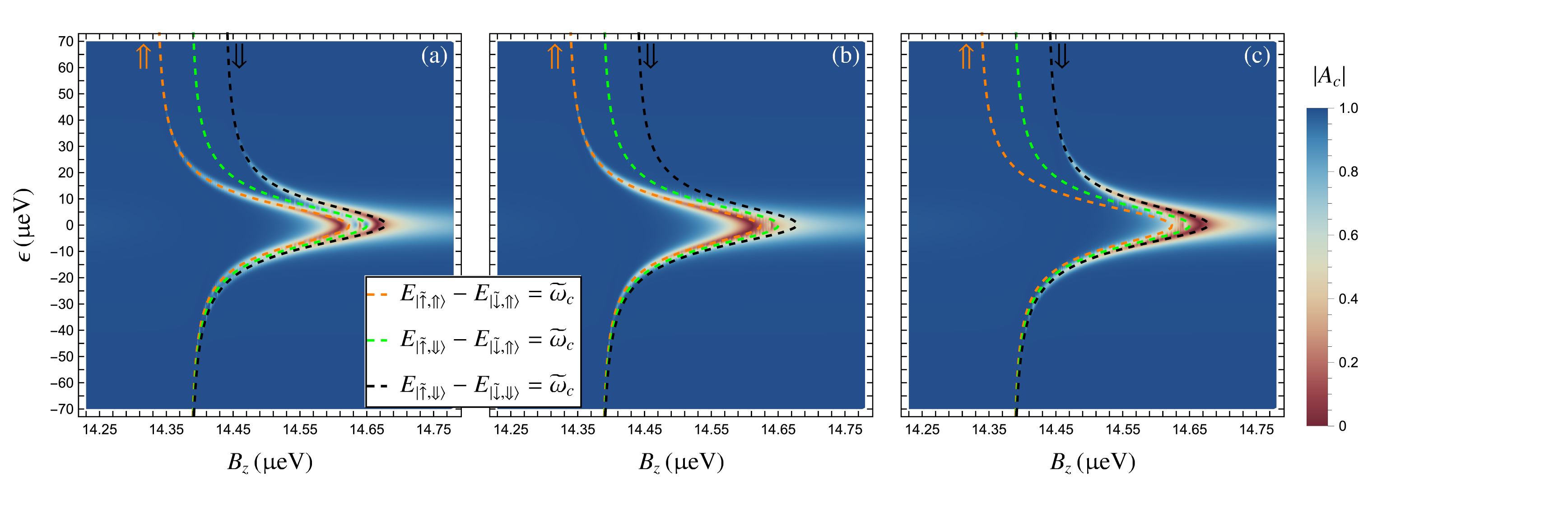}
	\caption{Cavity transmission $|A_c|$ on resonance, i.e., for $\omega=\omega_c$ with $\omega_c/2\pi=3.5\,\mathrm{GHz}= 14.5\,\upmu\mathrm{eV}$, as a function of the magnetic field $B_z$ and the QD-donor detuning $\epsilon$ for different populations of the energy levels with (a) both nuclear spin states  equally populated, (b)  the nuclear spin up state  populated, (c)  the nuclear spin down state  populated. The  dashed lines indicate the points in the $(\epsilon,B_z)$ parameter plane at which the transitions indicated in Fig.~\ref{fig:energyspectrumheffdiag} fulfill one of the resonance conditions \eqref{eq:resonance_spin_flip} or \eqref{eq:resonance_flipflop}. The remaining system parameters are	$t_c=9\,\upmu\mathrm{eV}$, $b_x=1.62\,\upmu\mathrm{eV}$, $g_c/2\pi=13\,\mathrm{MHz}$, ${2\kappa_1/2\pi=2\kappa_2/2\pi=\kappa/2\pi=1.3\,\mathrm{MHz}}$, and $A=25\,\mathrm{MHz}\approx 0.1\,\upmu\mathrm{eV}$. For the phonon induced decoherence entering the decoherence superoperator $\gamma$ we assume the parameters $d=37\,\mathrm{nm}$, $\omega_0/2\pi=30\,\mathrm{GHz}$, $c_b=4000\,\mathrm{m/s}$ and $J_0$ determined by $J(2t_c=5.4\,\mathrm{GHz},d=120\,\mathrm{nm})=35\,\mathrm{MHz}$. The strength of the quasistatic charge noise affecting the detuning paramter $\epsilon$ is chosen to be $\sigma_{\epsilon}=1\,\upmu\mathrm{eV}$.}
	\label{fig:cavityTransmission}
\end{figure*}
\begin{figure}[ht]
\includegraphics[width=\columnwidth]{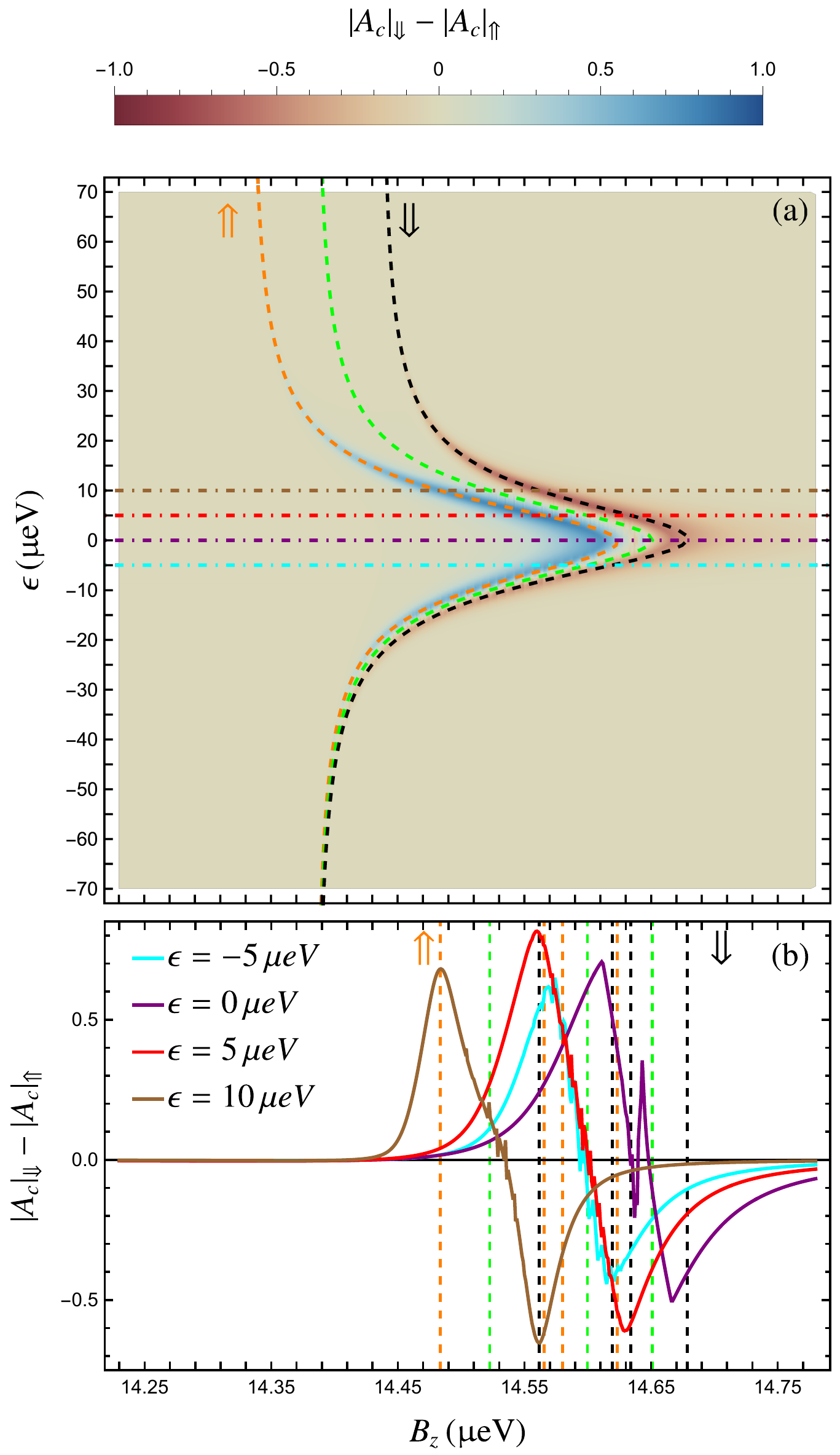}
    \caption{Difference of the cavity transmission $|A_c|$ between the nuclear spin down case  $|A_c|_{\Downarrow}$, and  the nuclear spin up case $|A_c|_{\Uparrow}$. The dot-dashed lines in (a) indicate the positions of the linecuts shown in (b). Parameters and dashed lines are as in Fig.~\ref{fig:cavityTransmission}. }
    \label{fig:readoutdifference}
\end{figure}
\begin{figure}[ht]
    \centering
    \includegraphics[width=\columnwidth]{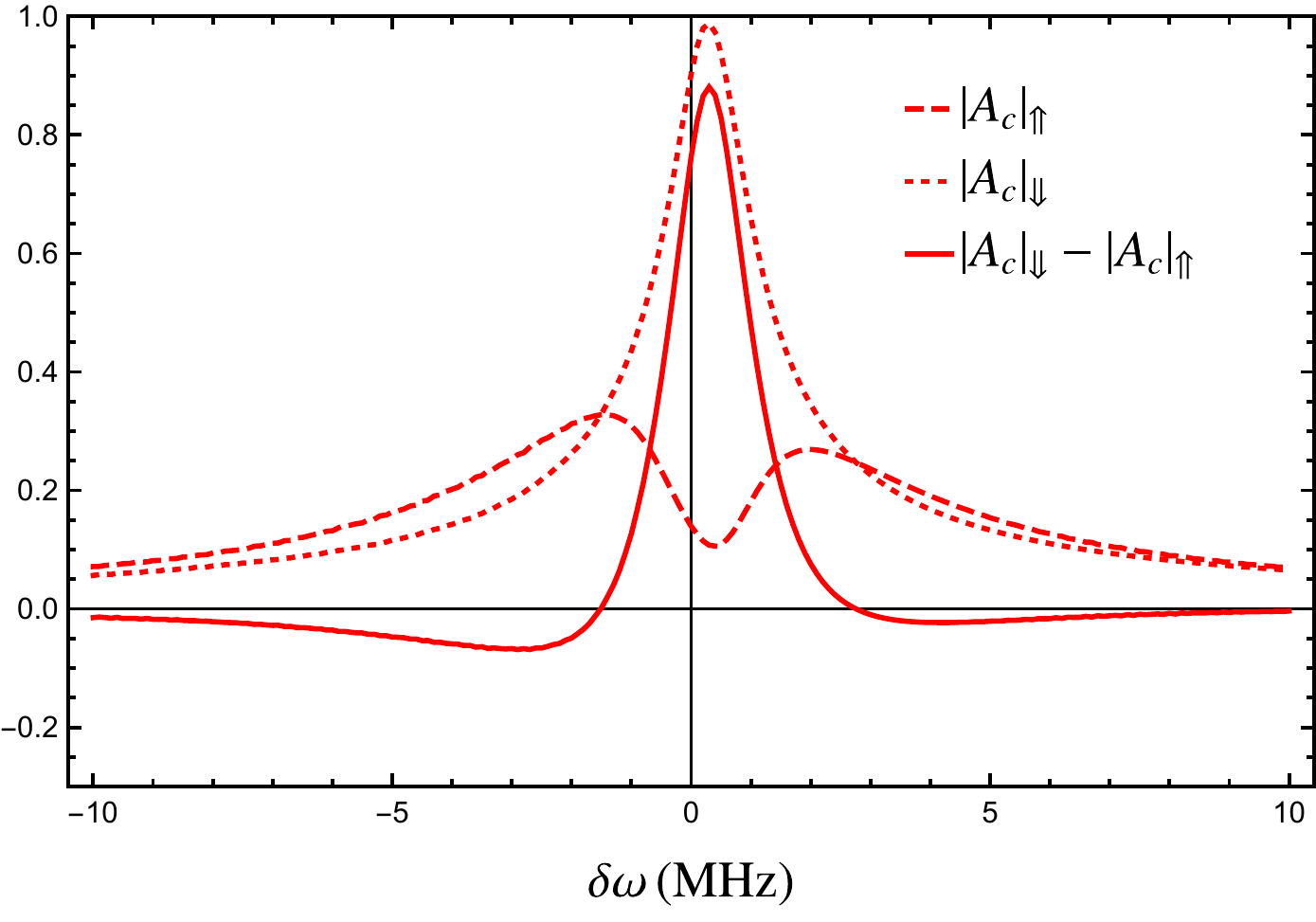}
    \caption{Cavity transmission $|A_c|_{\Uparrow}$, $|A_c|_{\Downarrow}$ and readout contrast, ${|A_c|_{\Downarrow}-|A_c|_{\Uparrow}}$, for $\epsilon=10\,\upmu\mathrm{eV}$ and $B_z=14.57\,\upmu\mathrm{eV}$, i.e. the point where the red and the second dashed orange line from the left in Fig.~\ref{fig:readoutdifference} intersect, as a function of the probe pulse detuning $\delta\omega$. The remaining parameters are as in Fig.~\ref{fig:cavityTransmission}.}
    \label{fig:deltaomega}
\end{figure}
The energy expectation values (\ref{eq:electrontransitionfrequency}) as a function of the QD-donor detuning $\epsilon$ and the various transition frequencies are presented in Fig.~\ref{fig:energyspectrumheffdiag}. 
Both Figure~\ref{fig:energyspectrumheffdiag} and Eq.~\eqref{eq:electrontransitionfrequency} show that the electron spin flip transition frequency depends on the state of the nuclear spin. More precisely, for a small electron spin mixing angle $\phi \ll 1$ the transition frequency with the nuclear spin in the states $\Uparrow$ and $\Downarrow$ differs by 
\begin{align}
\Delta=2\delta E_{\tilde{\sigma}}
\approx\frac{A}{2}\left(1+\sin\theta\right)+\frac{A^2\cos^2\theta\,\Omega}{8(\Omega^{2}-B_z^{2})}.
\label{eq:transitionfrequencydifference}
\end{align}
Hence, in the limits of large positive, zero, and large negative DQD detuning $\epsilon$, the shift in the resonance frequency $\Delta$ takes the values (note that $t_c>0$),
\begin{align}
\epsilon\gg 2t_c: & & \lim_{\theta\rightarrow \pi/2} \Delta\approx A, \nonumber\\
\epsilon=0:& & \Delta\approx \frac{A}{2}+\frac{A^2\Omega}{8(\Omega^{2}-B_z^{2})}, \\
\epsilon\ll - 2t_c: & & \lim_{\theta\rightarrow -\pi/2} \Delta\approx 0. \nonumber
\end{align}
The increasing impact of the nuclear spin on $\Delta$ with increasing QD-donor detuning is intuitively easy to understand: For $\epsilon\ll -2 t_c$ the electron is  localized in the left QD and therefore decoupled from the nuclear spin, at $\epsilon=0$ it is completely delocalized between the left QD and the donor, while it is trapped in the donor with a high probability for $\epsilon\gg 2 t_c$ such that the coupling to the nuclear spin is maximized. 

We note that in a DQD architecture with the second QD overlapping with an isoelectric $^{29}$Si nuclear spin, readout of the nuclear spin state has been realized for the maximal coupling scenario  $\epsilon\gg 2 t_c$ by probing the electron spin resonance frequency with frequency-selective ac magnetic field pulses \cite{hensen2020}. Even though the hyperfine interaction is as low as a few hundred kHz in such a device, we expect that alternatively our suggested readout method can be used as discussed in detail in Appendix \ref{appendix:hyperfineresolution}.

\section{Nuclear spin readout via the electron spin \label{sec:nuclearspinreadout}}

We now describe how the nuclear-spin dependent shift $\Delta$ of the electron-spin resonance frequency Eq.~\eqref{eq:transitionfrequencydifference} allows for a read-out of the nuclear spin.
The last term in \eqref{eq:Heffdiagonalmain} identifies the cavity resonance frequency including shifts of the empty cavity frequency $\omega_c$ due to the interaction with the QD-donor system. Thus, the cavity mode is resonant with the electron spin flip transition for a fixed nuclear spin state if
\begin{align}
E_{|\tilde{\uparrow},\Uparrow(\Downarrow)\rangle}-E_{|\tilde{\downarrow},\Uparrow(\Downarrow)\rangle}=\widetilde{\omega}_c,
\label{eq:resonance_spin_flip}
\end{align} and resonant with the electron spin-nuclear spin flip-flop transition if 
\begin{align}
E_{|\tilde{\uparrow},\Downarrow\rangle}-E_{|\tilde{\downarrow},\Uparrow\rangle}=\widetilde{\omega}_c.
\label{eq:resonance_flipflop}
\end{align}
 We expect a signature of the respective coupling in the cavity transmission in the vicinity of system parameters $\epsilon$, $t_c$, $B_z$, $b_x$, and $\omega_c$ for which one of these relations is fulfilled.
 
In order to verify our prediction we calculate the cavity transmission $A_c$ using input-output theory (Appendix~\ref{sec:InputOutput}) and compare the system parameters for which characteristic features emerge with those satisfying the resonance conditions derived above. The calculation of $A_c$ takes charge relaxation processes due to the phonon environment and quasi-static charge noise affecting the detuning parameter $\epsilon$ into account (see Appendix~\ref{appendix:DecoherenceModel} for details).
Figure~\ref{fig:cavityTransmission} shows the absolute value of the cavity transmission $|A_c|$ for three different populations of the hyperfine levels where (a) the two lowest energy levels are equally populated approximating the thermal equilibrium state for $T\gtrsim 30\,\mathrm{mK}$, i.e., the
QD-donor system is with equal probability in the states $|0\rangle$ and $|1\rangle$ which, up to small corrections, correspond to the nuclear spin up and down states 
$|\tilde{\downarrow},\Uparrow\rangle$ and $|\tilde{\downarrow},\Downarrow\rangle$, respectively; (b) only the ground state $\approx |\tilde{\downarrow},\Uparrow\rangle$ is populated; (c) only the excited state $\approx |\tilde{\downarrow},\Downarrow\rangle$ is populated. We point out that a single measurement will always be represented by the Figs.~\ref{fig:cavityTransmission}$(b)$ or $(c)$, while Fig.~\ref{fig:cavityTransmission}$(a)$ corresponds to the average over many measurements if the system is initialized with equal probability in the states $|0\rangle\approx |\tilde{\downarrow},\Uparrow\rangle$ and $|1\rangle \approx |\tilde{\downarrow},\Downarrow\rangle$ before the measurement. We find that the emerging characteristic features, given by a significantly reduced transmission due to the interaction of the cavity mode with the QD-donor system appear in the immediate vicinity of the parameters fulfilling the resonance conditions Eqns. \eqref{eq:resonance_spin_flip} and \eqref{eq:resonance_flipflop}, as indicated by the dashed lines in Fig.~\ref{fig:cavityTransmission}. 
One also observes that the signatures are less pronounced for $|\epsilon| \gg 2 t_c$. The last line of Eq.~\eqref{eq:V} shows that the electric dipole moment of the $|+\rangle\leftrightarrow|-\rangle$ transition is proportional to $\cos\theta$ and therefore decreases with increasing $|\epsilon/2t_c|$, which, in turn leads to the weakening of the effective couplings responsible for the observed signatures.

For the experimental realization of nuclear spin state readout it is essential to obtain a strong contrast between the signal for nuclear spin $\Uparrow$ and $\Downarrow$. In order to identify suitable readout points, we calculate the difference of the cavity transmission $|A_c|$ obtained for the excited state populated, $|A_c|_{\Downarrow}$, and the one with only the ground state populated, $|A_c|_{\Uparrow}$, i.e., Fig.~\ref{fig:cavityTransmission}$(b)$ is subtracted from Fig.~\ref{fig:cavityTransmission}$(c)$. The result presented in Fig.~\ref{fig:readoutdifference}$(a)$ unveils extended regions providing a high signal contrast for nuclear spin readout in the vicinity of the three resonance conditions, the two resonances \eqref{eq:resonance_spin_flip} and the resonance \eqref{eq:resonance_flipflop}, and weak QD-donor detuning in the range between $\epsilon=-10\,\upmu\mathrm{eV}$ and $\epsilon=15\,\upmu\mathrm{eV}$. The linecuts in Fig.~\ref{fig:readoutdifference}$(b)$ show that, within this range of QD-donor detuning, maximal contrast is achieved for points in the immediate vicinity of the resonance for nuclear spin $\Uparrow$. The amplitude difference of the readout contrast between the resonances for $\Uparrow$ and $\Downarrow$ can be attributed to a shift of the cavity resonance frequency caused by the interaction with the QD-donor system (see Appendix~\ref{appendix:hyperfineresolution} for more details).

Moreover, we can check the sensitivity of the readout contrast with respect to the cavity detuning from the probe field
\begin{align}
    \delta\omega=\omega-\omega_c,
\end{align}
for good readout points. To do so, we calculate 
$|A_c|_{\Uparrow}$, $|A_c|_{\Downarrow}$ and the readout contrast, $|A_c|_{\Downarrow}-|A_c|_{\Uparrow}$, for the point in Fig.~\ref{fig:readoutdifference} where the red and the second dashed orange line from the left intersect, as a function of the detuning $\delta\omega$. The result is presented in Fig.~\ref{fig:deltaomega} and shows a readout contrast larger than $0.2$ for $|\delta\omega|<1.5\,\mathrm{MHz}$. In addition, the figure allows one to identify the origin of reduced transmission in Fig.~\ref{fig:cavityTransmission}$(b)$ and the  resulting good readout contrast:
At the chosen readout point, the electron spin flip transition for nuclear spin $\Uparrow$ is close to resonance with the cavity, while the electron spin flip transition for nuclear spin  $\Downarrow$ is off-resonant. Due to the strong electron spin photon-coupling one observes Rabi splitting for nuclear spin $\Uparrow$ ($|A_c|_{\Uparrow}$ in Fig.~\ref{fig:deltaomega}), whereas $|A_c|_{\Downarrow}$ shows a single resonance located between the Rabi split modes of $|A_c|_{\Uparrow}$.  

To further characterize the nuclear spin measurement, we go beyond the input-output theory and  inspect Eqs.~\eqref{eq:Heffdiagonalmain} and \eqref{eq:Heintfull} describing the effective electronic Hamiltonian $H_{e,0}+H_{e,\mathrm{int}}$ in order to assess the expected measurement back-action. We note that since $[H_{e,0},\nu_z]=0$, the main part
of the hyperfine coupling leads to a nuclear spin readout in the form of a quantum non-demolition (QND) measurement \cite{braginsky1992}.  In general, $[H_{e,\mathrm{int}},\nu_z]\neq 0$ leading to small corrections to the QND behavior. However, for an adiabatic transfer of the electron from the left QD to the delocalized configuration between the QD and the donor and back under continuous transmission of a microwave field at constant frequency, we expect a recovery of the QND readout because the pure nuclear spin states are adiabatically transferred to eigenstates of $H_{e,0}+H_{e,\mathrm{int}}$.
Away from the resonance, the analogous argument holds with the non-resonant Hamiltonian \eqref{eq:effectiveHamiltonianTransmission}.

For the experimental verification of the suggested method for nuclear spin readout, we envision the following protocol: The cavity transmission is measured at one of the suitable readout points. Then, a nuclear spin resonance $\pi$-pulse is performed before the cavity transmission is probed again. Following the above discussion, successful nuclear spin readout is achieved if there is a significant difference in the absolute value of the transmission, and, depending on this value for the respective measurement, the state of the nuclear spin at the time of each measurement can be assigned.

\section{Nuclear spin photon coupling \label{sec:nuclear_spin-photon_coupling}}

 It has been shown that the nuclear spin of a QD-donor system can be controlled with a classical electric field \cite{boross2018}. However, this does not allow coherent information transfer between the nuclear spin and photons.
In order to assess the potential of the system for coherent coupling of the nuclear spin to cavity photons, we derive a Hamiltonian describing the effective dynamics of the nuclear spin interacting with the resonator mode while the remaining parts of the system are near the ground state.
More precisely, we investigate the dynamics of the subspace determined by the projection operator
\begin{align}
P_0=\frac{1-\tau_z}{2}\,\frac{1-\sigma_z}{2},
\end{align}
that defines the subspace spanned by the states ${|-,\downarrow,\Downarrow,n\rangle}$, 
${|-,\downarrow,\Uparrow,n\rangle}$, with $n=0,1,2,...$ .
 To do this, we apply a Schrieffer-Wolff transformation to decouple the subspaces defined by the projection operators $P_0$ and $Q_0=1-P_0$ \cite{bravyi2011}. Following the  procedure sketched in Sec. \ref{sec:nuclearspinreadout}, we determine $S_1$ and $S_2$, where $S_2$ is defined by 
$[H_0,S_2]=  -[P_0VP_0+Q_0VQ_0,S_1]$,
to obtain the effective Hamiltonian for the subspace defined by $P_0$ up to third order in the perturbation $V$ \cite{bravyi2011},
\begin{align}
H_{n}=&P_0H_0P_0+P_0VP_0+\tfrac{1}{2}P_0[S_1,P_0VQ_0+Q_0VP_0]P_0\nonumber\\
&+\tfrac{1}{2}P_0[S_2,P_0VQ_0+Q_0VP_0]P_0.
\label{eq:Heffformulanu}
\end{align}
In particular, the diagonal part of the effective Hamiltonian reads
\begin{align}
    H_{n,0}=&\frac{E_{\Bar{\nu}}}{2}\nu_z+\Bar{\omega}_ca^\dagger a+\frac{\delta E_{\Bar{\nu}}}{2}\nu_za^\dagger a,
\end{align}
with the expressions for ${E_{\Bar{\nu}},\,\Bar{\omega}_c}$ and $\delta E_{\Bar{\nu}}$ presented in Appendix \ref{appendix:EffectiveHnuclearspin}. We find that ${|\delta E_{\Bar{\nu}}|\ll |E_{\Bar{\nu}}|\ll \Bar{\omega}_c}$ if the electron is not entirely confined to the left QD. Thus, the microwave resonator and the donor nuclear spin flip transition cannot be tuned to resonance.
The coherent excitation exchange between these two subsystems is described by the term, 
\begin{align}
    g_{\nu}\left(\nu^{+}a^{\dagger}+\nu^{-}a\right),
    \label{eq:nuplusadager}
\end{align} 
within $H_n$ given in \eqref{eq:Heff}. We note that ${E_{|\Downarrow,n\rangle}>E_{|\Uparrow,n\rangle}}$ because ${E_{\Bar{\nu}}\approx-\tfrac{A}{4}(1+\sin(\theta))+\mathcal{O}(V^2)}$ such that Eq.~\eqref{eq:nuplusadager} is an excitation conserving interaction term.
The explicit form of the coupling constant $g_{\nu}$ in terms of the system parameters is given in Appendix~\ref{appendix:EffectiveHnuclearspin} and we find that nuclear spin to photon coupling strengths of  ${g_\nu\approx 0.5}$ MHz can be achieved. 
Given realistic values for the nuclear-spin 
and cavity loss rates, $\gamma\ll \kappa \approx 1$ MHz,
we note that the strong coupling regime for nuclear spin cavity QED ($g_\nu\gg \kappa, \gamma$) should be within reach. However, the coherent excitation exchange between these two subsystems is suppressed by the large detuning from resonance.

The raising or lowering of the nuclear spin state along with the creation or annihilation of a cavity photon results from the combined effect of the hyperfine interaction, the magnetic field gradient and the electric dipole interaction. 
The fundamental problem preventing resonant coupling is that, in principle, the energy splitting between the orbital states ($+$ and $-$) and the energy splitting between the electron spin states ($\uparrow$ and $\downarrow$) can simultaneously be tuned close to resonance with the microwave resonator, while, at the same time, the energy splitting between the nuclear spin states ($\Uparrow$ and $\Downarrow$) is far off-resonant because the nuclear gyromagnetic ratio is $\approx 1000$ times smaller than the one of the electron spin.

\section{Conclusion \label{sec:conclusion}}
In conclusion, we have investigated a system composed of a donor nuclear spin coupled to the spin of a single electron in a QD-donor architecture via the hyperfine interaction. The electron is subject to a homogeneous magnetic field and a magnetic field gradient perpendicular to the homogeneous component, while it is also dipole coupled to a microwave resonator.
We demonstrate that the effective excitation-conserving nuclear spin-photon interaction resulting from the combined effect of the hyperfine interaction, the electric dipole interaction, and the magnetic field gradient cannot directly be tuned to resonance. 

Nevertheless, we show that the signature of the strong electron spin-photon coupling \cite{benito2017} in the cavity transmission is altered due to the hyperfine interaction. We find well separated signatures for the electron spin-photon coupling with the nuclear spin in the states $\Uparrow$ and $\Downarrow$, whereby the splitting of the two signatures is determined by the hyperfine interaction strength $A$. For a $^{31}\mathrm{P}$ donor in the strained Si quantum well with $A\approx 25\,\mathrm{MHz}$ we expect that recent experimental setups are able to resolve the split signatures individually. Moreover, we identified good readout points at which one finds a high contrast between the measurement signal for the two opposing nuclear spin polarizations. Therefore, the cavity transmission allows for a readout of the nuclear spin state and for the measurement of the hyperfine interaction strength.

\section*{Acknowledgments}
We thank M\'{o}nica Benito and N. Tobias Jacobson for helpful discussions. This work has been supported by ARO grant number W911NF-15-1-0149.

\appendix
\section{ QD-donor system \label{appendix:QD-donor system}}
In this Appendix, we present a simulation of the QD-donor architecture that allows us to obtain a rough estimation for the size of the achievable tunnel coupling strength and the electric dipole moment. 

In Si/SiGe heterostructures electrons in the Si quantum well are strongly confined in growth direction, defining the vertical position of the QD in the Si quantum well \cite{zwanenburg2013}. Additional lateral confinement, required to form a QD, can be realized with a layer of gate electrodes a few tenths of nanometers above the quantum well. In order to obtain a lateral QD-donor architecture, a $^{31}$P donor has to be implanted in the quantum well. In the following we assume a separation of 56 nm between the gate layer and the plane containing the donor in the quantum well in line with recent Si/SiGe QD systems \cite{mi2017,mi2018,zajac2015}. 

\begin{figure}[ht]
    \centering
    \includegraphics[width=\columnwidth]{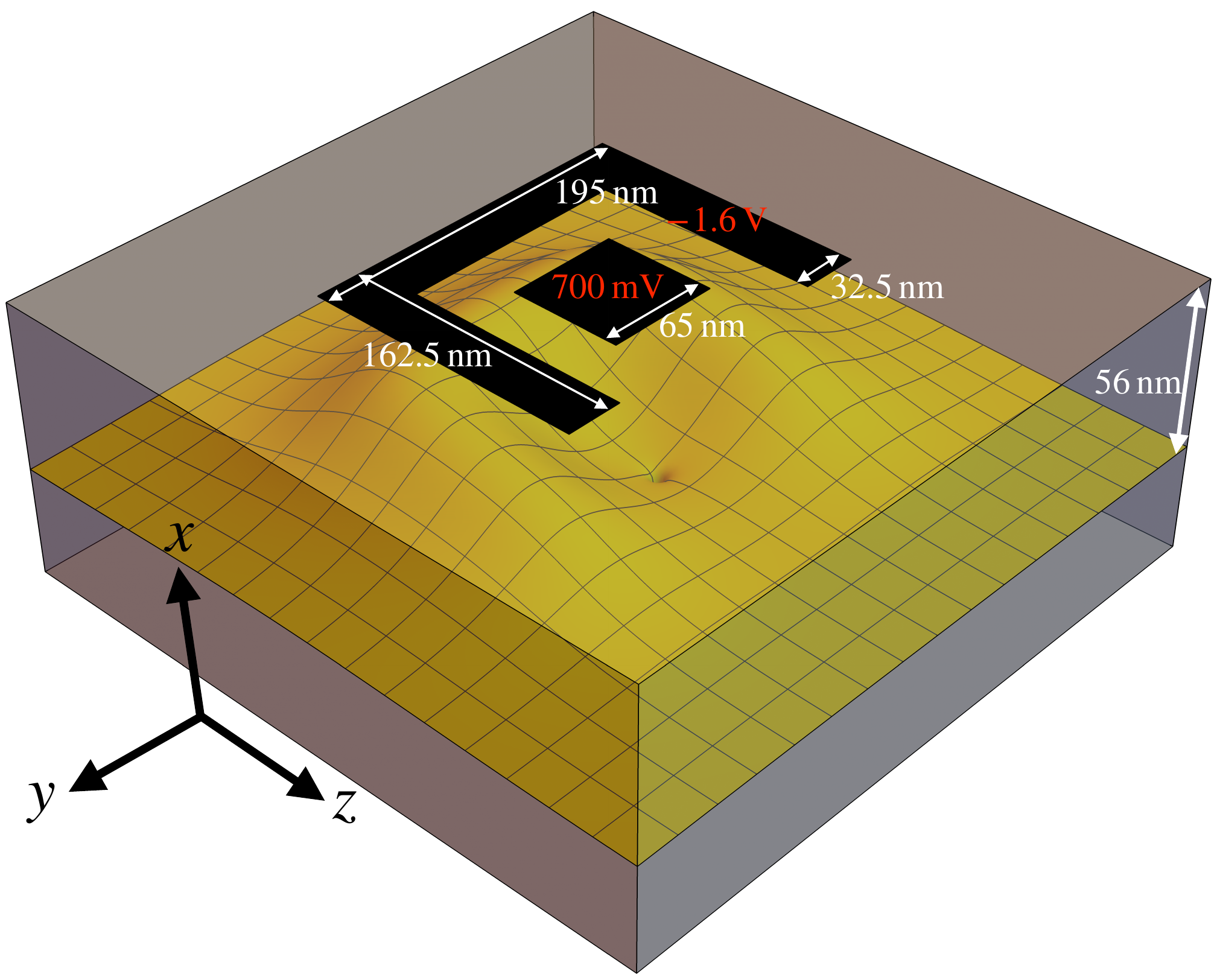}
    \caption{Surface gate architecture (black) and electron confinement potential $-e \Phi$ in the donor plane generated by the gates and the ionized donor. The gate voltages are chosen as indicated. The origin of the coordinate system is set below the center of the quadratic gate at the bottom of the gate layer and the axes are as shown in the figure. }
    \label{fig:gatesandpotential}
\end{figure}

As a first step, we determine the electrostatic potential $\Phi$ in the donor-plane generated by the gate architecture illustrated in  Fig.~\ref{fig:gatesandpotential} and the ionized donor by numerically solving the Poisson equation
\begin{align}
    \Delta \Phi=-\frac{1}{\epsilon_0\epsilon_r}\rho(\vec{r}),
\end{align}
with $\epsilon_r=11.7$ the relative permittivity of Si. The applied gate voltages are considered by setting the boundary conditions accordingly, while the ionized donor is modelled by the homogeneous spherical charge density
\begin{align}
    \rho(\vec{r})=
    \left\{
    \begin{array}{cc}
    \frac{e}{4/3 \pi r_{c}^3} & \left|\vec{r}-\vec{r}_{d}\right|\leq r_c \\
    0 & \left|\vec{r}-\vec{r}_{d}\right|>r_c
\end{array}
\right. ,
\end{align}
with $\vec{r}_{d}=(-56,0,30)\,\mathrm{nm}$ the donor position, i.e. the donor is implanted 56 nm below the gate layer and displaced by 30 nm in $z$-direction relative to the center of the rectangular gate in Fig.~\ref{fig:gatesandpotential}. We choose $r_c=0.95\,\mathrm{nm}$ ensuring that the correct $^{31}$P donor binding energy (45.5 meV) is achieved if no gate voltage is applied. The resulting electron confinement potential $-e\Phi$ for the gate voltages indicated in Fig.~\ref{fig:gatesandpotential} in the donor plane is also shown in Fig.~\ref{fig:gatesandpotential}. We note that our calculations do not consider layers of different materials and material interfaces between these layers present in real Si/SiGe devices. However, due to the similar dielectric constants of Si and Si$_{0.7}$Ge$_{0.3}$, the  resulting effects on the electrostatic potential in the donor plane are small and can be compensated by slightly modifying the gate architecture and the applied gate voltages. 

In the following, the level detuning between the lowest-lying QD and donor state is adjusted by an external electric field in $z$-direction. Alternatively, the level detuning could also be controlled with more complex gate architectures. 

Given the strong confinement in growth direction, it suffices to solve the two-dimensional Schr\"odinger equation for an estimation of the QD-donor tunnel coupling strength $t_c$. Explicitely, the Schr\"odinger equation reads
\begin{align}
    0=&-\frac{\hbar^2}{2 m_{\bot}}\left(\partial_y^2+\partial_z^2\right)\psi(y,z)\nonumber\\
    &-e (\Phi(y,z)+ E_{\mathrm{ext}} z)\psi(y,z),
    \label{eq:Schroedinger}
\end{align}
with $m_{\bot}=0.192\,m_e$ the transverse effective electron mass in Si grown along $[100]$ and $E_{\mathrm{ext}}$ the electric field strength.

\begin{figure}[ht]
\centering 
\includegraphics[width=\columnwidth]{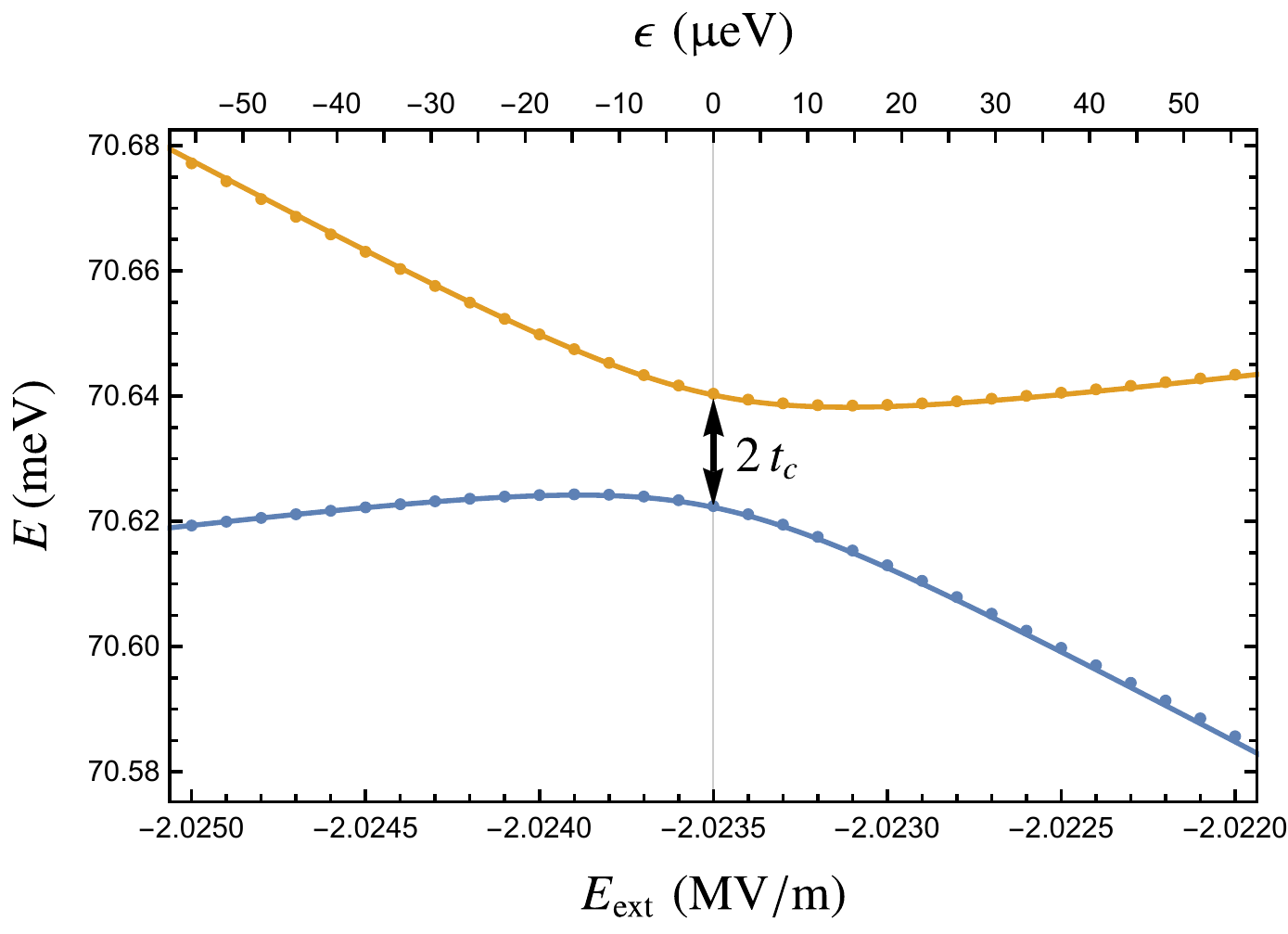}
\caption{Energies of the ground state and the first excited state of the QD-donor system as a function of the external electric field $E_{\mathrm{ext}}$ determining the level detuning $\epsilon$. The dots are obtained from numerical solutions of \eqref{eq:Schroedinger}, while the solid lines describe a simplified two-level system.}
\label{fig:avoidedcrossing}
\end{figure}

The energies of the ground  and  first excited states of the QD-donor system as a function of the external electric field obtained by numerically solving \eqref{eq:Schroedinger} are shown as the points in Fig.~\ref{fig:avoidedcrossing}. The spectrum shows an avoided crossing at $E^0_{\mathrm{ext}}\approx -2.0235\,\mathrm{MV/m}$ with minimal energy difference $\Delta E_{\mathrm{min}}\approx 18\,\upmu\mathrm{eV}$. We find good agreement between the simulation (points) and a simplified two level model (solid lines) with tunnel coupling $2 t_c=\Delta E_{\mathrm{min}}=18\,\upmu\mathrm{eV}$ and level detuning $\epsilon=-e\left(E^0_{\mathrm{ext}}-E_{\mathrm{ext}}\right) d$, where $d=37\,\mathrm{nm}$ gives the QD-donor distance discussed later. This observation justifies the orbital two-level model in \eqref{eq:H0tilde} and shows that a sizeable tunnel coupling strength is reachable in lateral QD-donor devices despite the sharp confinement potential of the donor. 

\begin{figure}[ht]
\centering 
\includegraphics[width=\columnwidth]{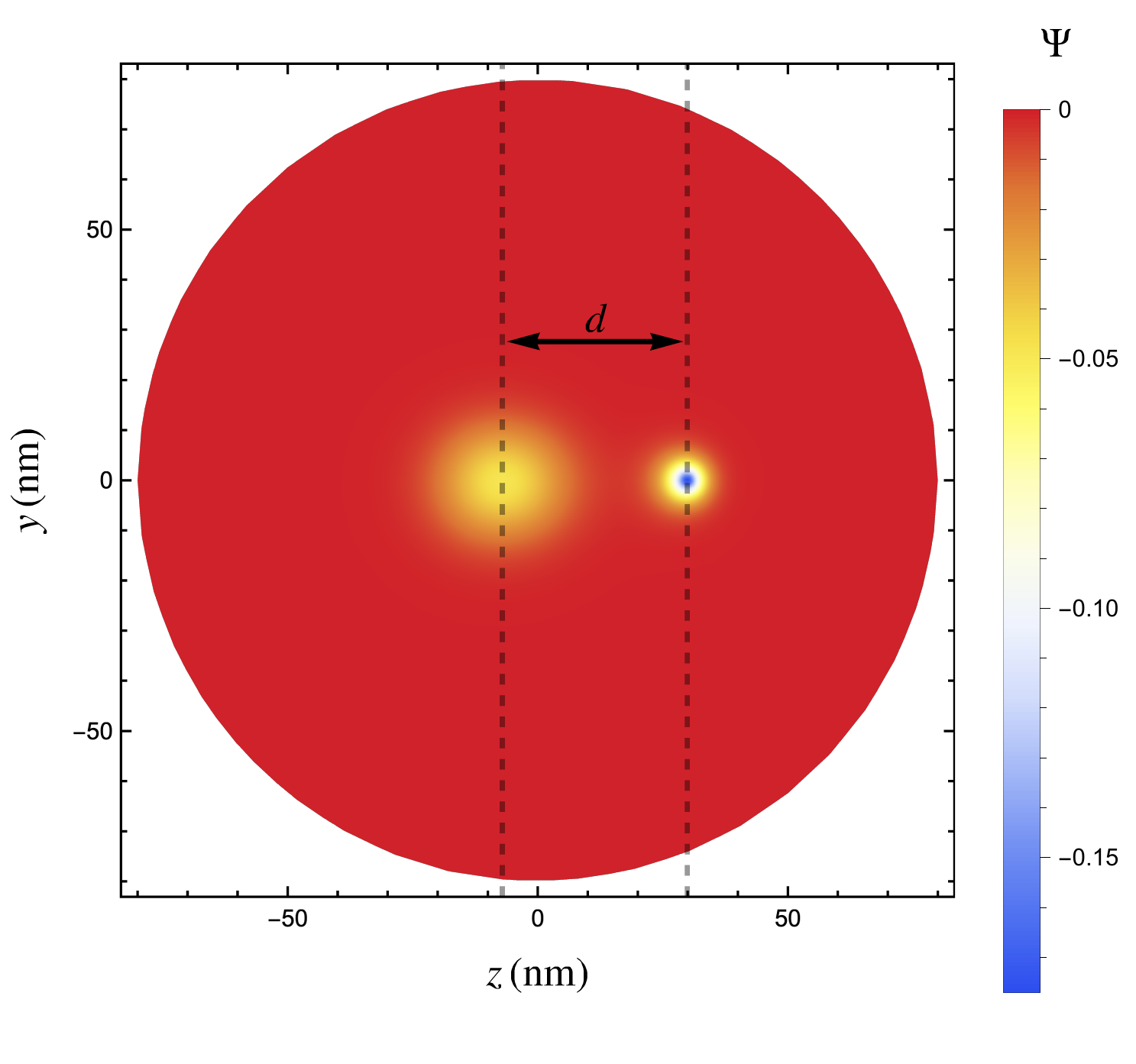}
\caption{Ground state wave function in the donor plane for $E_{\mathrm{ext}}=E^0_{\mathrm{ext}}$.}
\label{fig:groundstatewavefunction}
\end{figure}

For the suggested nuclear spin readout method a notable tunnel coupling strength alone is not sufficient, since also the charge-photon coupling  $g_c$ has to be sufficiently strong. The charge-photon coupling strength depends linearly on the electric dipole moment $e d$, where $d$ is the QD-donor distance \cite{burkard2020}. For the setup discussed in this section, one can extract $d\approx 37\,\mathrm{nm}$ from Fig.~\ref{fig:groundstatewavefunction}, that shows the ground state wave function at $E_{\mathrm{ext}}=E^0_{\mathrm{ext}}$, where the electron equally populates QD and donor. At other values of $E_{\mathrm{ext}}$ in the range given in Fig.~\ref{fig:avoidedcrossing}, the wave functions of the ground state and the first excited state have to be compared, but similar results for $d$ are obtained. In DQD devices typical values for the inter dot distance are $100-120\,\mathrm{nm}$. Therefore, the charge-photon coupling strength in the QD-donor device is expected to be $\approx 1/3$ of the coupling strength reported for DQD devices.

\section{Effective Hamiltonian of the strong electron spin-photon coupling \label{appendix:effective_Hamiltonian}}
The Schrieffer Wolff transformation \eqref{eq:Heffeformula} yields the Hamiltonian
\begin{align}
H_{e}=&(\alpha_1+\alpha_2 \nu_z)\sigma_z+\alpha_{3}\nu_z+(\alpha_{4}+\alpha_{5}\sigma_z\nu_z)a^{\dagger}a\nonumber\\
&+(\alpha_{6}+\alpha_{7}\nu_z)(\sigma^++\sigma^-)\nonumber\\
&+(\alpha_{8}+\alpha_{9}\sigma_z)(\nu^++\nu^-)\nonumber\\
& +\alpha_{10}(2 \sigma^+\nu^-+2\sigma^-\nu^+)\nonumber\\
&+(\alpha_{11}+\alpha_{12}\sigma_z\nu_z)(a+a^{\dagger})\nonumber\\
&+\alpha_{13}(2 \sigma^+\nu^-+2\sigma^-\nu^+)(a+a^{\dagger})\nonumber\\
&+(\alpha_{14}+(\alpha_{15}-i \alpha_{16})\nu_z )(\sigma^+a+\sigma^-a^{\dagger})\nonumber\\
&+(\alpha_{14}+(\alpha_{15}+i \alpha_{16})\nu_z )(\sigma^+a^{\dagger}+\sigma^-a)\nonumber\\
&+(\alpha_{17}+\alpha_{18}\sigma_z\nu_z)(a^2+(a^{\dagger})^2),
\label{eq:effectiveHamiltonianTransmission}
\end{align}
with the coefficients $\alpha_1$ to $\alpha_{18}$ discussed below. 
The transformation requires the coupling between states of the subspace defined by $P_0$ and $Q_0$ to be much smaller than the energy separation of those states \cite{bravyi2011}. 
In the present case this requirement is ensured provided that the following relations hold:
\begin{align}
&g_c |\cos\theta|\ll\min\left( \left|-\Omega-\frac{A}{4}\sin\theta+\omega_c\right|,\right.\nonumber\\
 &\left.\,\,\,\,\left|-\Omega+\frac{A}{4}|\sin\theta|-\omega_c\right|, \left|-\Omega+\frac{A}{4}\sin\theta+\omega_c\right|\right),\\[10pt]
&\frac{A}{8}|\cos\theta|\ll\left|-\Omega+\frac{A}{4}\left|\sin\theta\right|\right|,\\[10pt]
&\frac{b_x}{2}|\cos\theta|\ll \min\left(\left|-\Omega-B_z-\frac{A}{4}\right|,\left|-\Omega-B_z+\frac{A}{4}\right|,\right.
\nonumber\\
 & \left.\,\,\,\,\left|-\Omega+B_z+\frac{A}{4}\right|,\left|-\Omega+B_z-\frac{A}{4}\right|\right).
\end{align}
However, we consider a regime where ${\Omega, B_z ,\omega_c\gg  A,b_x,g_c}$. Hence we can use the approximations
\begin{align}
&-\Omega\pm\frac{A}{4}\sin\theta\varpm \omega_c\approx-\Omega\varpm \omega_c,\\
&-\Omega\pm\frac{A}{4}\sin\theta\approx -\Omega,\\
&-\Omega\varpm B_z\pm\frac{A}{4}\approx-\Omega\varpm B_z.
\end{align}
Taking into account the above approximations, the parameters of the Hamiltonian \eqref{eq:effectiveHamiltonianTransmission} read

\begin{align}
\alpha_{1}=&\frac{B_z}{2}-\left(\frac{A^2}{32}+\frac{b_x^2}{4}\right)\cos^2\theta\,\frac{B_z}{\Omega^2-B_z^2},\\
\alpha_{2}=&\frac{A}{8}(1+\sin\theta)+\frac{A^2\cos^2\theta}{32}\frac{\Omega}{\Omega^2-B_z^2},\\
\alpha_{3}=&\frac{A^2\cos^2\theta}{32}\frac{B_z}{\Omega^2-B_z^2},\\
\alpha_4=&\omega_c-2g_c^2\cos^2\theta\frac{\Omega}{\Omega^2-\omega_c^2},\\
\alpha_5=&0,\\
\alpha_{6}=&-\frac{b_x\sin\theta}{2},\\
\alpha_{7}=&-\frac{Ab_x\cos\theta}{16}\frac{B_z}{\Omega^2-B_z^2},\\
\alpha_{8}=&\frac{Ab_x\cos^2\theta}{8}\frac{B_z}{\Omega^2-B_z^2},\\
\alpha_{9}=&\frac{Ab_x\cos^2\theta}{8}\frac{B_z}{\Omega^2-B_z^2},
\end{align}
\begin{align}
\alpha_{10}=&\frac{A}{8}(1+\sin\theta)+\frac{A^2\cos^2\theta}{64}\left(\frac{1}{\Omega}+\frac{\Omega}{\Omega^2-B_z^2}\right),\\
\alpha_{11}=&-g_c\sin\theta,\\
\alpha_{12}=&\frac{A g_c \cos^2\theta}{8}\left(\frac{1}{\Omega}+\frac{\Omega}{\Omega^2-\omega_c^2}\right),\\
\alpha_{13}=&\frac{Ag_c\cos^2\theta}{8}\left(\frac{\Omega}{\Omega^2-B_z^2}+\frac{\Omega}{\Omega^2-\omega_c^2}\right),\\
\alpha_{14}=&-\frac{b_xg_c\cos^2\theta}{2}\left(\frac{\Omega}{\Omega^2-B_z^2}+\frac{\Omega}{\Omega^2-\omega_c^2}\right),\\
\alpha_{15}=&0,\\
\alpha_{16}=&0,\\
\alpha_{17}=&-g_c^2\cos^2\theta\,\frac{\Omega}{\Omega^2-\omega_c^2},\\
\alpha_{18}=&0.
\end{align}


We note, that the term proportional to $\alpha_6$ causing a mixing between the electron spin states is not not negligible. Thus, we need to account for this term when calculating transition energies between states that we expect to resemble the actual eigenstates of the Hamiltonian. To this end, we transform $H_{\mathrm{eff}}$ into the eigenbasis of 
\begin{align}
\alpha_1\sigma_z+\alpha_{6}(\sigma^++\sigma^-).
\end{align}
The transformed basis states are
\begin{align}
|-,\tilde{\downarrow},\Uparrow(\Downarrow)\rangle=&\cos\frac{\phi}{2}|-,\downarrow,\Uparrow(\Downarrow)\rangle\nonumber\\
&-\sin\frac{\phi}{2}|-,\uparrow,\Uparrow(\Downarrow)\rangle,\\
|-,\tilde{\uparrow},\Uparrow(\Downarrow)\rangle=&\sin\frac{\phi}{2}|-,\downarrow,\Uparrow(\Downarrow)\rangle\nonumber\\
&+\cos\frac{\phi}{2}|-,\uparrow,\Uparrow(\Downarrow)\rangle,
\end{align}
with the electron spin mixing angle 
\begin{align}
\phi=\arctan\left(\frac{\alpha_6}{\alpha_1}\right).
\end{align}
Since, here, the magnetic field gradient $b_x$ is small compared to the homogeneous magnetic field $B_z$ one finds ${|\alpha_6|\ll |\alpha_{1}|}$, such that the electron spin mixing angle is small and therefore the states $|\tilde{\downarrow}(\tilde{\uparrow})\rangle$ are predominantly the electron spin states $|\downarrow(\uparrow)\rangle$ up to small contributions of the opposite electron spin state. The electron spin Pauli operators transform as 
\begin{align}
\sigma_x\rightarrow&\sin\phi\,\widetilde{\sigma}_z+\frac{\alpha_6}{|\alpha_6|}\cos\phi\,\widetilde{\sigma}_x,\\
\sigma_y\rightarrow&\frac{\alpha_6}{|\alpha_6|}\widetilde{\sigma}_y,\\
\sigma_z\rightarrow&\cos\phi\,\widetilde{\sigma}_z-\frac{\alpha_6}{|\alpha_6|}\sin\phi\,\widetilde{\sigma}_x,
\end{align}
with the Pauli operators $\widetilde{\sigma}_i$ operating on the $|\tilde{\downarrow}(\tilde{\uparrow})\rangle$ states. We divide the transformed Hamiltonian $\widetilde{H}_{e}$ in a diagonal part $\widetilde{H}_{e,0}$ and a part containing the interactions between the basis states $\widetilde{H}_{e,\mathrm{int}}$. For $\widetilde{H}_{e,0}$ we find
\begin{align}
\widetilde{H}_{e,0}=&\left(\frac{E_{\tilde{\sigma}}}{2}+\frac{\delta E_{\tilde{\sigma}}}{2}\nu_z\right)\tilde{\sigma}_z+ \frac{E_{\nu}}{2}\nu_z+\widetilde{\omega}_c a^{\dagger}a,
\label{eq:Heffdiagonal}
\end{align}
with
\begin{align}
    &E_{\tilde{\sigma}}=2\sqrt{\alpha_1^2+\alpha_6^2},\\
    &\delta E_{\tilde{\sigma}}=2(\alpha_2 \cos\phi+\alpha_7\sin\phi),\\
    &E_{\nu}=2\alpha_3,\\
    &\widetilde{\omega}_c=\alpha_4.
\end{align}
Since we consider a parameter regime with ${B_z,\omega_c\gg A,b_x,g_c}$ we find ${ E_{\tilde{\sigma}},\widetilde{\omega}_c\gg E_{\nu},\delta E_{\tilde{\sigma}}}$. If we additionally assume the effective cavity frequency $\widetilde{\omega}_c$ to be close to resonance with the electron spin transition frequency $E_{\tilde{\sigma}}$, we can apply the RWA to $H_{e,\mathrm{int}}$ keeping terms rotating with frequencies $\ll E_{\tilde{\sigma}}$:
\begin{align}
H_{e,\mathrm{int}}=&\left(\alpha_{8}\left[1+\cos\phi\,\tilde{\sigma}_z\right]+\alpha_{10}\sin\phi\,\tilde{\sigma}_z\right)\left(\nu^+ +\nu^-\right)\nonumber\\
&- g_{\tilde{\sigma}\nu}\sin^2\frac{\phi}{2}\left(\tilde{\sigma}^+ \nu^+ a +\tilde{\sigma}^- \nu^- a^{\dagger}\right)\nonumber\\
&+g_{\tilde{\sigma}\nu}\cos^2\frac{\phi}{2}\left(\tilde{\sigma}^+ \nu^- a +\tilde{\sigma}^- \nu^+ a^{\dagger}\right)\nonumber\\
&+\left(g_{\tilde{\sigma}}\cos\phi+\delta g_{\tilde{\sigma}}\sin\phi\,\nu_z\right)\left(\tilde{\sigma}^+a+\tilde{\sigma}^- a^{\dagger}\right),\label{eq:Heintfull}
\end{align}
with
\begin{align}
    &g_{\tilde{\sigma}\nu}=2\frac{\alpha_6}{|\alpha_6|}\alpha_{13}=-\mathrm{sgn}\,\epsilon\,2\alpha_{13},\\
    &g_{\tilde{\sigma}}=\frac{\alpha_6}{|\alpha_6|}\alpha_{14}=-\mathrm{sgn}\,\epsilon\,\alpha_{14},\\
    &\delta g_{\tilde{\sigma}}=\mathrm{sgn}\,\epsilon\,\alpha_{12},
\end{align}
and
\begin{align}
    g_{\tilde{\sigma}}>\delta g_{\tilde{\sigma}}.
\end{align}

We note that corrections to next higher order of the energy level structure of the reduced effective Hamiltonian due to the upper orbital subspace originate from combined effects involving a transition from the lower orbital subspace to the upper orbital subspace, a transition from the upper orbital subspace to the lower orbital subspace and a transition in either the lower or the upper orbital subspace \cite{winkler2003}. In this work we are mainly focused on determining transition frequencies from \eqref{eq:Heffdiagonal}. Thus, the relevant third order corrections are those contributing to $\alpha_1,\,\alpha_2,\,\alpha_3,\,\alpha_4,\,\alpha_6$ and $\alpha_7$. Given \eqref{eq:V}, it is straightforward to show that there are no third order corrections to $\alpha_1,\,\alpha_2,\,\alpha_3$ and $\alpha_4$.  
For the given Hamiltonian the third order corrections to spin flips, i.e. corrections to $\alpha_6$ and $\alpha_7$ have an upper bound given by $\delta^{(3)}\approx \frac{(b_x/2)^3}{(\Omega-B_z)\Omega}$, provided that $B_z \approx \omega_c$ and that $b_x/2 > A/4,\, g_c$. 
Using the parameters from the main text, $b_x=1.62\,\upmu\mathrm{eV}$, $\Omega\approx 18\, \upmu\mathrm{eV}$ and $B_z\approx 14.5 \,\upmu\mathrm{eV}$, one finds $|(E_{\tilde{\sigma}}(\alpha_1,\alpha_6+\delta^{(3)})-E_{\tilde{\sigma}}(\alpha_1,\alpha_6))/E_{\tilde{\sigma}}(\alpha_1,\alpha_6)|<6\cdot 10^{-4}$ and $|(\delta E_{\tilde{\sigma}}(\alpha_1,\alpha_6+\delta^{(3)},\alpha_7+\delta^{(3)})-\delta E_{\tilde{\sigma}}(\alpha_1,\alpha_6,\alpha_7))/ \delta E_{\tilde{\sigma}}(\alpha_1,\alpha_6,\alpha_7)|<6\cdot 10^{-3}$. Hence, third and higher order corrections due to the Schrieffer Wolff transformation can be neglected in the calculation of the transition frequencies.

\section{Input-Output Theory \label{sec:InputOutput}}
To investigate the transmission through the cavity interacting with the QD-donor system we use input-output theory.
We divide the Hamiltonian into three parts
\begin{align}
\widetilde{H}=\widetilde{H}_{\mathrm{sys}}+\widetilde{H}_I+\widetilde{H}_{\mathrm{cav}},
\label{eq:HtildeInoutOutput}
\end{align}
with the system Hamiltonian, $\widetilde{H}_{\mathrm{sys}}=\widetilde{H}_0+\widetilde{H}_{e-n}$, comprising the single electron in the QD-donor confinement potential \eqref{eq:H0tilde} and its hyperfine interaction to the nuclear spin \eqref{eq:He-n}. The eigenstates and the corresponding eigenenergies of $\widetilde{H}_{\mathrm{sys}}$ are denoted $|n\rangle$ and $E_n$ with $E_n\leq E_{n+1}$, respectively.
In the eigenbasis of $\widetilde{H}_{\mathrm{sys}}$ Eq.~\eqref{eq:HtildeInoutOutput} reads
\begin{align}
H=\sum_{n}E_n \sigma_{nn}+\omega_c a^{\dagger}a +g_c\sum_{m,n}d_{mn}\sigma_{mn}\left(a+a^{\dagger}\right),
\label{eq:H}
\end{align}
where the eigenstates of $\widetilde{H}_{\mathrm{sys}}$ define the operators $\sigma_{nm}=|n\rangle \langle m|$.
For the quantum Langevin equations for $\dot{a}(t)$ and $\dot{\sigma}_{nm}(t)$ one obtains
\begin{align}
\dot{a}&=i \left[H,a\right]-\frac{\kappa}{2}a+\sum_{i=1,2}\sqrt{\kappa_i}a_{\mathrm{in},i},\\
\dot{\sigma}_{mn}&= i \left[H,\sigma_{mn}\right]-\sum_{m'n'}\gamma_{mn,m'n'}\sigma_{m'n'}
+\sqrt{2\gamma}\mathcal{F},
\end{align}
with $a_{\mathrm{in},1}(t)$ and $a_{\mathrm{in},2}(t)$ the incoming parts of the external fields at the cavity ports 1 and 2. Moreover we have introduced the decoherence superoperator with matrix elements $\gamma_{mn,m'n'}$, that is discussed in detail in Appendix~\ref{appendix:DecoherenceModel}, and the quantum noise $\mathcal{F}$ of the QD-donor system. In the following discussion we will neglect the quantum noise $\mathcal{F}$.
Using Eq.~(\ref{eq:H}), we find
\begin{align}
\dot{a}&=-i \omega_c a-i g_c\sum_{m,n}d_{mn}\sigma_{mn}-\frac{\kappa}{2}a\nonumber\\
&\,\,\,\,\,\,\,+\sqrt{\kappa_1}a_{\mathrm{in},1}+\sqrt{\kappa_2}a_{\mathrm{in},2},
\label{eq:adot(t)}
\end{align}

and
\begin{align}
\dot{\sigma}_{mn}(t)=& i(E_m-E_n)\sigma_{mn}+ig_c\left(\sum_{m'}d_{m'm}\sigma_{m'n}\right.\nonumber\\
&\left.-\sum_{n'}d_{nn'}\sigma_{mn'}\right)\left(a+a^{\dagger}\right)\nonumber\\
&-\sum_{m',n'}\gamma_{mn,m'n'}\sigma_{m'n'}+\sqrt{2\gamma}\mathcal{F}.
\end{align}

We will now decompose $\sigma_{mn}(t)$ into a contribution independent of the cavity coupling $g_c$ and a part that is linear in $g_c$, while higher order contributions in $g_c$ are neglected,
\begin{align}
\sigma_{mn}(t)\approx \sigma_{mn}^{(0)}(t)+\sigma_{mn}^{(1)}(t)+\mathcal{O}(g_c^2),
\label{eq:sigmanmcontributions}
\end{align}
where $\sigma_{mn}^{(0)}(t)$ and $\sigma_{mn}^{(0)}(t)$ are zeroth and first order in $g_c$, respectively.
We replace the operators $\sigma_{mn}^{(0)}$ in \eqref{eq:sigmanmcontributions} by their expectation values 
\begin{align}
\langle\sigma_{mn}^{(0)}\rangle_{t}=\delta_{mn} p_m,
\label{eq:sigmanm0}
\end{align}
where $p_m$ are the average populations of the energy levels obtained for $g_c=0$.
Following the above discussion one obtains 
\begin{align}
\frac{\mathrm{d}}{\mathrm{d}t}\langle\sigma_{mn}^{(1)}\rangle_t=&i(E_m-E_n)\langle\sigma_{mn}^{(1)}\rangle_t\nonumber\\
&+ig_c d_{nm} \left( p_n -p_m\right)\left(\langle a\rangle_t+\langle a\rangle_t^*\right)\nonumber\\
&-\sum_{m',n'}\gamma_{mn,m'n'}\langle\sigma^{(1)}_{m'n'}\rangle_t.
\end{align}
for the expectation values of the operators first order in $g_c$.
A Fourier transformation to frequency space yields
\begin{align}
\langle\sigma_{mn}^{(1)}\rangle_{\omega}=&\frac{g_c d_{nm}\left(p_n-p_m\right)}{\left(E_n-E_m\right)-\omega-i\gamma_{mn,mn}}\left(\langle a\rangle_{\omega}+\langle a\rangle_{-\omega}^*\right)\nonumber\\
&+i\frac{\sum_{m',n'}\left(1-\delta_{mm'}\delta_{nn'}\right)\gamma_{mn,m'n'}\langle\sigma^{(1)}_{m'n'}\rangle_{\omega}}{\left(E_n-E_m\right)-\omega-i\gamma_{mn,mn}}.
\label{eq:deltasigmanmlineareqautions}
\end{align}
If the cavity has a large quality factor ${\mathcal{Q}=\omega_c/\kappa\gg 1}$ and is probed close to resonance such that ${|\omega-\omega_c|\ll \omega_c}$ a RWA for the cavity mode can be applied showing that the impact of ${\langle a \rangle_{-\omega}^*}$ is negligible \cite{kohler2018}. In this operating regime we can solve the set of linear equations \eqref{eq:deltasigmanmlineareqautions} to obtain the susceptibilities $\chi_{mn}(\omega)$,
\begin{align}
\langle\sigma_{mn}^{(1)}\rangle_{\omega}=\chi_{mn}(\omega)\langle a\rangle_{\omega}.
\label{eq:susceptibility}
\end{align}
Calculating the expectation value of \eqref{eq:adot(t)}, considering \eqref{eq:sigmanmcontributions} as well as \eqref{eq:sigmanm0}, before employing a Fourier transform to frequency space and using \eqref{eq:susceptibility} yields
\begin{align}
\langle a\rangle_{\omega}=&\frac{\sqrt{\kappa_1}\langle a_{\mathrm{in},1} \rangle_{\omega}+\sqrt{\kappa_2}\langle a_{\mathrm{in},2}\rangle_{\omega}}{i\left(\omega_c-\omega\right)+ig_c\sum_{m,n}\chi_{mn}(\omega)d_{mn}+\kappa/2}\nonumber\\
&-\frac{ig_c\sum_{n}d_{nn}p_n\delta(\omega)}{i\left(\omega_c-\omega\right)+ig_c\sum_{m,n}\chi_{mn}(\omega)d_{mn}+\kappa/2}.
\end{align}
According to input-output theory, the incoming and outgoing fields are related by \cite{gardiner1985}
\begin{align}
a_{\mathrm{out},\nu}-a_{\mathrm{in},\nu}=\sqrt{\kappa_{\nu}}a.
\end{align}
We consider a scenario with $\langle a_{\mathrm{in,2}} \rangle=0$
and can finally calculate the cavity transmission
\begin{align}
A_c&=\frac{\langle a_{\mathrm{out},2}\rangle_{\omega}}{\langle a_{\mathrm{in},1}\rangle_{\omega}}=\frac{\sqrt{\kappa_2}\langle a\rangle_{\omega}}{\langle a_{\mathrm{in,1}}\rangle_{\omega}}\nonumber\\
&=\frac{-i\sqrt{\kappa_1\kappa_2}}{\left(\omega_c-\omega\right)+g_c\sum_{m,n}\chi_{mn}(\omega)d_{mn}-i\kappa/2}.
\end{align}

\section{Noise \label{appendix:DecoherenceModel}}
The dominant noise effects in the present system arise due to charge relaxation processes induced by the phonon environment and due to charge noise. 

\subsection{Charge relaxation due to the phonon environment}
The electron phonon interaction for an electron in a QD-donor system is described by the Hamiltonian 
\begin{align}
	\widetilde{H}_{\mathrm{e-ph}}=\sum_{\mathbf{q},\nu}\lambda_{\mathbf{q}\nu}\widetilde{\tau}_z(a^{\dagger}_{\mathbf{q}\nu}+a_{\mathbf{q}\nu}),
\end{align}
with the momentum $\mathbf{q}$ and mode $\nu$ dependent coupling constants $\lambda_{\mathbf{q}\nu}$, and the corresponding phonon creation and annihilation operators. Let us recall that $\widetilde{\tau}_z$ transforms to $\sum_{m,n}d_{mn}\sigma_{mn}$
 under the transformation to the eigenbasis of $\widetilde{H}_{\mathrm{sys}}$. Hence, using Fermi's golden rule we find the transition rate from eigenstate $|n\rangle$ to $|m\rangle$ at zero temperature
\begin{align}
\gamma_{n\rightarrow m}=&2\pi\sum_{\mathbf{q},\nu}\left|\langle \mathbf{q},\nu|\langle m| \sum_{\mathbf{q}',\nu'}\lambda_{\mathbf{q}',\nu'}\left(\sum_{m',n'} d_{m'n'}\sigma_{m'n'}\right)\right.\nonumber\\
&\left.\left(c_{\mathbf{q}'\nu'}+c_{\mathbf{q}'\nu'}^{\dagger}\right)|n\rangle |0\rangle \right|^2\delta \left((\mathrm{E}_n-\mathrm{E}_m)-\varepsilon_{\mathbf{q},\nu}\right)\nonumber \\
=&2\pi\sum_{\mathbf{q},\nu}\left|\lambda_{\mathbf{q},\nu}\right|^2 \delta \left((\mathrm{E}_n-\mathrm{E}_m)-\varepsilon_{\mathbf{q},\nu}\right) \left|d_{mn}\right|^2, \nonumber\\
=&2\pi J(E_n-E_m)\left|d_{mn}\right|^2,
\label{eq:Gammamn}
\end{align}
where $|0\rangle$ denotes the phonon vacuum and $|\mathbf{q},\nu\rangle$ is a single phonon state with energy $\varepsilon_{\mathbf{q},\nu}$. ${J(\nu)=\sum_{\mathbf{q},\nu}\left|\lambda_{\mathbf{q}',\nu'}\right|^2 \delta \left(\nu-\varepsilon_{\mathbf{q},\nu}\right)}$ is the phonon spectral density. \\
We can also calculate the transition rate for the orbital transition $|+\rangle\rightarrow |-\rangle $ for $\epsilon=0$:
\begin{align}
\gamma_{+\rightarrow -}=2\pi J(\Omega).
\label{eq:Gamma+-}
\end{align}
This relation allows one to specify the scale factor $J_0$ introduced below in \eqref{eq:phononspectraldensity} to describe the phonon spectral density, because values for this rate were reported in a recent experiment \cite{mi2018} considering a similar setup. \\
Due to the inversion symmetry of the unit cell of the crystal structure of silicon electron phonon coupling is caused by bulk deformation potential coupling \cite{yu2010} and the phonon spectral density at low energy can be modeled by \cite{gullans2018,brandes2005}
\begin{align}
J(\nu)=J_0\left(\frac{\nu}{\omega_0}\right)^3\left(1-\mathrm{sinc}\left(\frac{\nu d}{c_b}\right)\right)e^{-\frac{\nu^2}{2\omega_0^2}},
\label{eq:phononspectraldensity}
\end{align}
where $J_0$ is a scale factor, $\omega_0$ a cutoff frequency, $d$ the spacing between the QD and the donor and $c_b$ the speed of sound. \\ 
To capture the decoherence effects due to the phonon environment we use a markovian quantum master equation in Lindblad form with the jump operator \cite{nathan2020}
\begin{align}
L=\sum_{m,n}\sqrt{2\pi J(E_n-E_m)}d_{mn}\sigma_{mn}.
\end{align}
We assume the phonon bath to be at zero temperature such that only transitions to lower energy states are possible, i.e.,
\begin{align}
L&=\sum_{m<n}\sqrt{2\pi J(E_n-E_m)}d_{mn}\sigma_{mn}\nonumber\\
&=\sum_{m<n}j_{mn}d_{mn}\sigma_{mn},
\end{align}
with $j_{mn}=\sqrt{2\pi J(E_n-E_m)}$. \\
One can calculate the mean value for the decoherence dynamics of the operators $\sigma_{mn}$ to identify the elements $\gamma_{mn,m'n'}$ of the decoherence superoperator:
\begin{align}
\frac{\mathrm{d}}{\mathrm{d}t}&\langle\sigma_{mn} \rangle_t=\mathrm{Tr}\left\{\sigma_{mn}\mathcal{D}[L]\rho(t)\right\}\nonumber\\
=&\sum_{m',n'}\left[\left(j_{mm'}\right)^*j_{nn'}-\frac{1}{2}\left\{\delta_{mm'}\left(\sum_{k}\left(j_{kn}\right)^*j_{kn'}\right)\right.\right.\nonumber\\
&\left.\left.+\delta_{nn'}\left(\sum_{k}\left(j_{km'}\right)^*j_{km}\right)\right\}\right]\langle\sigma_{m'n'}\rangle_t\nonumber\\
=&-\sum_{m',n'}\gamma_{mn,m'n'}\langle\sigma_{m'n'}\rangle_t,
\end{align} 
where $\mathcal{D}[L]$ represents the Dissipator superoperator $\mathcal{D}[L]\rho(t)=L\rho(t)L^{\dagger}-\frac{1}{2}\left(\rho(t)L^{\dagger}L+L^{\dagger}L\rho(t)\right)$ \cite{breuer2007}.

\subsection{Charge noise}
In semiconductor QD architectures charge noise is omnipresent. Charge noise leads to fluctuations of the electrostatic potentials in the proximity of the QD and the donor. Hence, charge noise mainly affects the QD-donor system in the form of fluctuations of the detuning parameter $\epsilon\rightarrow \epsilon+\delta\epsilon$. Here, quasistatic and gaussian distributed fluctuations of $\epsilon$ with standard deviation $\sigma_{\epsilon}$ are considered. In this context quasistatic means that $\delta\epsilon$ does not change during a single run of the experiment, but differs for different runs, wherefore we include the noise in our calculation of a quantity by convolving the respective quantity with the gaussian distribution. In particular one has
\begin{align}
A_c(\epsilon)=\frac{1}{\sqrt{2\pi\sigma_{\epsilon}^2}}\int_{-\infty}^{\infty}A_c(\epsilon')\,e^{-\frac{(\epsilon-\epsilon')^2}{2\sigma_{\epsilon}^2}}\mathrm{d}\epsilon'.
\end{align}

\section{Characteristics of the readout contrast \label{appendix:hyperfineresolution}}
\begin{widetext}
In order to derive an expression estimating the readout contrast, we use the derived effective Hamiltonian (\eqref{eq:Heffdiagonalmain} and \eqref{eq:HeffeintRWA}) for input-output theory. Following the steps outlined in Appendix \ref{sec:InputOutput}, one finds  
\begin{align}
\dot{a}&=-i \widetilde{\omega}_c a-i \left(g_{\tilde{\sigma}}\cos\phi+\delta g_{\tilde{\sigma}}\sin\phi\,\nu_z\right)\widetilde{\sigma}^- -ig_{\widetilde{\sigma}\nu}\cos^2\frac{\phi}{2}\tilde{\sigma}^- \nu^+ -\frac{\kappa}{2}a
+\sqrt{\kappa_1}a_{\mathrm{in},1}+\sqrt{\kappa_2}a_{\mathrm{in},2},
\end{align}
where we have neglected the contribution from the first term in \eqref{eq:HeffeintRWA} because $\sin^2 \frac{\phi}{2}\ll 1$. Moreover, straightforward calculations result in
\begin{align}
    \frac{\mathrm{d}}{\mathrm{d}t}\left(|\tilde{\downarrow},\Uparrow(\Downarrow)\rangle \langle |\tilde{\uparrow},\Uparrow(\Downarrow)|\right)=&i\left(E_{|\tilde{\downarrow},\Uparrow(\Downarrow)\rangle}-E_{|\tilde{\uparrow},\Uparrow(\Downarrow)\rangle}\right)\left(|\tilde{\downarrow},\Uparrow(\Downarrow)\rangle \langle \tilde{\uparrow},\Uparrow(\Downarrow)|\right)
    \nonumber\\
    &+i\left(g_{\tilde{\sigma}}\cos\phi\varpm\delta g_{\tilde{\sigma}}\sin\phi\right)\left(|\tilde{\uparrow},\Uparrow(\Downarrow)\rangle\langle\tilde{\uparrow},\Uparrow(\Downarrow)|-|\tilde{\downarrow},\Uparrow(\Downarrow)\rangle\langle\tilde{\downarrow},\Uparrow(\Downarrow)|\right)a,
\end{align}
and
\begin{align}
    \frac{\mathrm{d}}{\mathrm{d}t}\left(|\tilde{\downarrow},\Uparrow\rangle \langle |\tilde{\uparrow},\Downarrow|\right)=&i\left(E_{|\tilde{\downarrow},\Uparrow\rangle}-E_{|\tilde{\uparrow},\Downarrow\rangle}\right)\left(|\tilde{\downarrow},\Uparrow\rangle \langle \tilde{\uparrow},\Downarrow|\right)
   +ig_{\tilde{\sigma}\nu}\cos^2\frac{\phi}{2}\left(|\tilde{\uparrow},\Downarrow\rangle\langle\tilde{\uparrow},\Downarrow|-|\tilde{\downarrow},\Uparrow\rangle\langle\tilde{\downarrow},\Uparrow|\right)a,
\end{align}
where in comparison to the discussion in Appendix~\ref{sec:InputOutput} the ideal decoherence free scenario is considered for simplicity. In analogy to Appendix~\ref{sec:InputOutput}, the susceptiblities for the three different processes can be determined:
\begin{align}
   \bigg\langle\left(|\tilde{\downarrow},\Uparrow(\Downarrow)\rangle \langle \tilde{\uparrow},\Uparrow(\Downarrow)|\right)\bigg\rangle_{\omega}=
    \frac{\left(g_{\tilde{\sigma}}\cos\phi\varpm \delta g_{\tilde{\sigma}}\sin\phi\right)\left(p_{|\tilde{\uparrow},\Uparrow(\Downarrow)\rangle}-p_{|\tilde{\downarrow},\Uparrow(\Downarrow)\rangle}\right)}{\left(E_{|\tilde{\uparrow},\Uparrow(\Downarrow)\rangle}-E_{|\tilde{\downarrow},\Uparrow(\Downarrow)\rangle}\right)-\omega}\langle a\rangle_{\omega}=\chi_{\Uparrow (\Downarrow)}(\omega)\langle a\rangle_{\omega},
    \label{eq:chiupanddown}
\end{align}
and
\begin{align}
   \bigg\langle\left(|\tilde{\downarrow},\Uparrow\rangle \langle \tilde{\uparrow},\Downarrow|\right)\bigg\rangle_{\omega}=
    \frac{g_{\tilde{\sigma}\nu}\cos^2\frac{\phi}{2}\left(p_{|\tilde{\uparrow},\Downarrow\rangle}-p_{|\tilde{\downarrow},\Uparrow\rangle}\right)}{\left(E_{|\tilde{\uparrow},\Downarrow\rangle}-E_{|\tilde{\downarrow},\Uparrow\rangle}\right)-\omega}\langle a\rangle_{\omega}=\chi_{\Uparrow \Downarrow}(\omega)\langle a\rangle_{\omega}.
\end{align}
With the susceptibilities, one obtains 
\begin{align}
\langle a\rangle_{\omega}=&\frac{\sqrt{\kappa_1}\langle a_{\mathrm{in},1} \rangle_{\omega}+\sqrt{\kappa_2}\langle a_{\mathrm{in},2}\rangle_{\omega}}{i\left(\widetilde{\omega}_c-\omega\right)+i\left[\left(g_{\tilde{\sigma}}\cos\phi+\delta g_{\tilde{\sigma}}\sin\phi\right)\chi_{\Uparrow}(\omega)+\left(g_{\tilde{\sigma}}\cos\phi-\delta g_{\tilde{\sigma}}\sin\phi\right)\chi_{\Downarrow}(\omega)\right]+ig_{\tilde{\sigma}\nu}\cos^2\frac{\phi}{2} \chi_{\Uparrow\Downarrow}(\omega)+\kappa/2}
\end{align}
and therefore the cavity transmission reads
\begin{align}
A_c&=\frac{\langle a_{\mathrm{out},2}\rangle_{\omega}}{\langle a_{\mathrm{in},1}\rangle_{\omega}}=\frac{\sqrt{\kappa_2}\langle a\rangle_{\omega}}{\langle a_{\mathrm{in,1}}\rangle_{\omega}}\nonumber\\
&=\frac{-i\sqrt{\kappa_1\kappa_2}}{\left(\widetilde{\omega}_c-\omega\right)+\left(g_{\tilde{\sigma}}\cos\phi+\delta g_{\tilde{\sigma}}\sin\phi\right)\chi_{\Uparrow}(\omega)+\left(g_{\tilde{\sigma}}\cos\phi-\delta g_{\tilde{\sigma}}\sin\phi\right)\chi_{\Downarrow}(\omega)+g_{\tilde{\sigma}\nu}\cos^2\frac{\phi}{2} \chi_{\Uparrow\Downarrow}(\omega)-i\kappa/2}.
\label{eq:AcHeff}
\end{align}
Using the explicit expressions for the susceptibilities, the terms in the denominator can be expressed as
$\left(g_{\tilde{\sigma}}\cos\phi\varpm\delta g_{\tilde{\sigma}}\sin\phi\right)\chi_{\Uparrow(\Downarrow)}\propto\frac{\left(g_{\tilde{\sigma}}\cos\phi\pm\delta g_{\tilde{\sigma}}\sin\phi\right)^2}{\left(E_{|\tilde{\uparrow},\Uparrow(\Downarrow)\rangle}-E_{|\tilde{\downarrow},\Uparrow(\Downarrow)\rangle}\right)-\omega}$ and $g_{\tilde{\sigma}\nu}\cos^2\frac{\phi}{2}\,\chi_{\Uparrow\Downarrow}\propto\frac{\left(g_{\tilde{\sigma}\nu}\cos^2\frac{\phi}{2}\right)^2}{\left(E_{|\tilde{\uparrow},\Downarrow\rangle}-E_{|\tilde{\downarrow},\Uparrow\rangle}\right)-\omega}$. Since $\left(g_{\tilde{\sigma}}\cos\phi\pm\delta g_{\tilde{\sigma}}\sin\phi\right)^2>\left(g_{\tilde{\sigma}\nu}\cos\frac{\phi}{2}\right)^2$, the term $\propto \chi_{\Uparrow\Downarrow}$ in the denominator leads to a sharp feature in the transmission that does not significantly influence the readout contrast away from this feature, and is therefore neglected in the following. Equation \eqref{eq:chiupanddown} shows that $\chi_{\Downarrow}=0\,(\chi_{\Uparrow}=0)$ if the system is initially prepared in the state characterized by $p_{|\tilde{\downarrow},\Uparrow\rangle}=1\,(p_{|\tilde{\downarrow},\Downarrow\rangle}=1)$. Probing the cavity at its resonance frequency ($\omega=\omega_c$) and approximating  $\tilde{\omega}_c\approx\omega$ allows one to omit the first term in the denominator of \eqref{eq:AcHeff}. Taking into account all these considerations and assuming $\kappa_1=\kappa_2=\kappa/2$, one finds
\begin{align}
1-|A_c|_{\Uparrow(\Downarrow)}^2=\frac{\left(\frac{2g_{\Uparrow (\Downarrow)}^2}{\kappa}\right)^2}{\left(\frac{2g_{\Uparrow (\Downarrow)}^2}{\kappa}\right)^2+\left[\omega-\left(E_{|\tilde{\uparrow},\Uparrow(\Downarrow)\rangle}-E_{|\tilde{\downarrow},\Uparrow(\Downarrow)\rangle}\right)\right]^2},
\label{eq:Lorentzian}
\end{align}
with 
$g_{\Uparrow(\Downarrow)}=g_{\tilde{\sigma}}\cos\phi\varpm \delta g_{\tilde{\sigma}}\sin\phi$.
In the parameter domains suggested for nuclear spin readout with $\left|\frac{\Omega}{\Omega^2-B_z^2}\right|$ and $b_x\ll Bz$, the couplings $g_{\Uparrow(\Downarrow)}$ can be considered constant over extended domains in $B_z$, while the detuning $\epsilon$ is fixed. Moreover, $E_{|\tilde{\uparrow},\Uparrow(\Downarrow)\rangle}-E_{|\tilde{\downarrow},\Uparrow(\Downarrow)\rangle}$ is approximately linear in $B_z$. Thus, Eq. \eqref{eq:Lorentzian} describes a Lorentzian line shape with maximum value 1 as a function of $B_z$ for fixed detuning $\epsilon$ with width $2\left(\frac{2g_{\Uparrow (\Downarrow)}^2}{\kappa}\right)$. From Eq. \eqref{eq:Lorentzian}, we find 
\begin{align}
    1-|A_c|_{\Uparrow(\Downarrow)}=1-\sqrt{1-\frac{\left(\frac{2g_{\Uparrow (\Downarrow)}^2}{\kappa}\right)^2}{\left(\frac{2g_{\Uparrow (\Downarrow)}^2}{\kappa}\right)^2+\left[\omega-\left(E_{|\tilde{\uparrow},\Uparrow(\Downarrow)\rangle}-E_{|\tilde{\downarrow},\Uparrow(\Downarrow)\rangle}\right)\right]^2}},
    \label{eq:oneminusAc}
\end{align}
again describing a line shape with maximum value 1 and symmetric around the resonance defined by {$\omega-\left(E_{|\tilde{\uparrow},\Uparrow(\Downarrow)\rangle}-E_{|\tilde{\downarrow},\Uparrow(\Downarrow)\rangle}\right)$}. For values of $B_z$ separated from the resonance by $\xi \left(\frac{2g_{\Uparrow (\Downarrow)}^2}{\kappa}\right)$, i.e. {$\omega-\left(E_{|\tilde{\uparrow},\Uparrow(\Downarrow)\rangle}-E_{|\tilde{\downarrow},\Uparrow(\Downarrow)\rangle}\right)=\xi \left(\frac{2g_{\Uparrow (\Downarrow)}^2}{\kappa}\right)$}, one has 
\begin{align}
    1-|A_c|_{\Uparrow(\Downarrow)}=1-\sqrt{\frac{\xi^2}{1+\xi^2}}. 
\end{align}
In the suggested nuclear spin readout method, the discrimination between $\Uparrow$ and $\Downarrow$ is based on the transmission difference for the two nuclear spin states. The signal shapes for $\Uparrow$ and $\Downarrow$ are almost similar in parameter domains with $b_x\ll B_z$ such that $g_{\Uparrow}\approx g_{\Downarrow}\approx g_{\tilde{\sigma}}$, while the maxima of $1-|A_c|_{\Uparrow}$ and $1-|A_c|_{\Downarrow}$ are separated by $\Delta=2\delta E_{\tilde{\sigma}}$. Therefore, given the line shape \eqref{eq:oneminusAc}, the absolute value of the readout contrast $|A_c|_{\Downarrow}-|A_c|_{\Uparrow}$ is maximal for values of $B_z$ where $1-|A_c|_{\Uparrow(\Downarrow)}=1$, while 
$1-|A_c|_{\Downarrow(\Uparrow)}=1-\sqrt{\frac{\xi_{\Delta}^2}{1+\xi_{\Delta}^2}}$ with $\xi_{\Delta}=\frac{\Delta}{\left(\frac{2g_{\tilde{\sigma}}^2}{\kappa}\right)}=\frac{\delta E_{\tilde{\sigma}}\kappa}{g_{\tilde{\sigma}}^2}$, such that 
\begin{align}
   \left|\left(|A_c|_{\Downarrow}-|A_c|_{\Uparrow}\right)\right|=\sqrt{\frac{\left(\frac{\delta E_{\tilde{\sigma}}\kappa} {g_{\tilde{\sigma}}^2}\right)^2}{1+\left(\frac{\delta E_{\tilde{\sigma}}\kappa} {g_{\tilde{\sigma}}^2}\right)^2}},
\end{align}
where $B_z\approx \omega_c$ can be chosen to determine $g_{\tilde{\sigma}}$ and $\delta E_{\tilde{\sigma}}$.

However, this result does not account for the small but finite detuning $\widetilde{\omega}_c-\omega$ if $\omega=\omega_c$ and the noise processes discussed in Appendix~\ref{appendix:DecoherenceModel}.  The detuning can be considered by the replacement 
$\omega-\left(E_{|\tilde{\uparrow},\Uparrow(\Downarrow)\rangle}-E_{|\tilde{\downarrow},\Uparrow(\Downarrow)\rangle}\right) \rightarrow \frac{g_{\tilde{\sigma}}^2\left[\omega-\left(E_{|\tilde{\uparrow},\Uparrow(\Downarrow)\rangle}-E_{|\tilde{\downarrow},\Uparrow(\Downarrow)\rangle}\right)\right]}{\left(\widetilde{\omega}_c-\omega\right)\left[\omega-\left(E_{|\tilde{\uparrow},\Uparrow(\Downarrow)\rangle}-E_{|\tilde{\downarrow},\Uparrow(\Downarrow)\rangle}\right)\right]+g_{\tilde{\sigma}}^2}$
in the Eqs.~\eqref{eq:Lorentzian} and \eqref{eq:oneminusAc}, whereby the position of the respective maximum is not changed. 
Apart from that, we find for values of $B_z$ separated by $\delta B_z$ from the resonance, i.e. $\delta B_z=-\left[\omega-\left(E_{|\tilde{\uparrow},\Uparrow(\Downarrow)\rangle}-E_{|\tilde{\downarrow},\Uparrow(\Downarrow)\rangle}\right)\right]$, and $\widetilde{\omega}_c-\omega<0$ (the relevant case for $\omega=\omega_c$ and $2t_c>\omega_c$):
\begin{align}
   \left(\widetilde{\omega}_c-\omega\right)<0\wedge \delta B_z<0:\quad&\Bigg| \frac{g_{\tilde{\sigma}}^2(-\delta B_z)}{\left(\widetilde{\omega}_c-\omega\right)(-\delta B_z)+g_{\tilde{\sigma}}^2}\Bigg|>|\delta B_z|\quad \mathrm{if } \left(\widetilde{\omega}_c-\omega\right)\delta B_z<g_{\tilde{\sigma}}^2\\
   \left(\widetilde{\omega}_c-\omega\right)<0\wedge (-\delta B_z)>0:\quad&\Bigg| \frac{g^2_{\Uparrow(\Downarrow)}(-\delta B_z)}{\left(\widetilde{\omega}_c-\omega\right)(-\delta B_z)+g_{\tilde{\sigma}}^2}\Bigg|<|\delta B_z|\nonumber.
\end{align}
This observation implies that the side of the peak of $1-|A_c|$ with $\delta B_z>0$ decreases more slowly while the side with $\delta B_z<0$ decreases faster as a function of $B_z$ compared to the non-detuned scenario. The resonance for $\Uparrow$ is achieved for lower values of $B_z$ than the one for $\Downarrow$, and, therefore, the readout contrast at the resonance for $\Uparrow$ is determined by the fast decreasing flank of $1-|A_c|_{\Downarrow}$ at $\delta B_z=-\Delta$, while the readout contrast at the resonance condition for $\Downarrow$ is determined by the slow decreasing flank of $1-|A_c|_{\Uparrow}$ at $\delta B_z=\Delta$, wherefore the absolute value of the readout contrast is larger at the resonance for $\Uparrow$. This is exactly the behaviour of the line cuts shown in Fig.~\ref{fig:readoutdifference}(b) and is also visible in the Figs.~\ref{fig:readoutdifference}(a) and \ref{fig:readoutcontrasthyperfine}. In the same way as before we can define $\xi_{\Delta}^{\Uparrow(\Downarrow)}=\frac{\left(\frac{\varmp g_{\tilde{\sigma}}^2\Delta}{\varmp\left(\widetilde{\omega}_c-\omega\right)\Delta+g_{\tilde{\sigma}}^2}\right)}{\left(\frac{g_{\tilde{\sigma}}^2}{\kappa}\right)}=\frac{\varmp \kappa \Delta}{2\left[\varmp\left(\widetilde{\omega}_c-\omega\right)\Delta+g_{\tilde{\sigma}}^2\right]}$ and subsequently calculate the absolute value of the readout contrast at the resonance for $\Uparrow (\Downarrow)$:
\begin{align}
   \left|\left(|A_c|_{\Downarrow}-|A_c|_{\Uparrow}\right)\right|^{\Uparrow(\Downarrow)}=\sqrt{\frac{\left(\xi_{\Delta}^{\Uparrow(\Downarrow)}\right)^2}{1+\left(\xi_{\Delta}^{\Uparrow(\Downarrow)}\right)^2}}.
\end{align}
This result is in good agreement with the cut for $\epsilon=0$ (purple line) in Fig.~\ref{fig:readoutdifference}(b). However, it overestimates the extremal values of the cuts for $\epsilon>0$ significantly, because there, the $B_z$ values at which the extremal readout contrast occurs, are sensitive to small changes in the detuning $\epsilon$ (see also Figs.~\ref{fig:cavityTransmission} and \ref{fig:readoutdifference}(a)). Thus, the quasistatic charge noise considered in the Figures (for details see Appendix \ref{appendix:DecoherenceModel}) reduces the absolute value of the extremal readout contrast.

The readout contrast observed in Fig.~\ref{fig:readoutdifference} is certainly sufficiently large for nuclear spin readout in recent experimental devices. Nevertheless, we can comment on the minimal hyperfine interaction strength leading to a sufficient contrast for readout. Using \eqref{eq:AcHeff}, one can numerically calculate the absolute value of the readout contrast and account for quasistatic charge noise in the way discussed in Appendix~\ref{appendix:DecoherenceModel}. A map of the readout contrast dependence on the hyperfine interaction strength is presented in Fig.~\ref{fig:readoutcontrasthyperfine}. The plots clearly show that there are readout points with $\big|\left(|A_c|_{\Downarrow}-|A_c|_{\Uparrow}\right)\big|>0.01$ in domains with $A<1\,\mathrm{MHz}$. This is sufficient for readout because recent cQED experiments are able to measure $|A_c|/|A_0|$ with precision of fractions of a percent \cite{hartke2018}. Therefore, we expect that the suggested nuclear spin readout technique is also applicable in a DQD system with an isoelectric nuclear spin, e.g. $^{29}$Si, at the position of one of the QDs because $A$ in the range of several hundred kHz is reported for such devices \cite{hensen2020}.

\begin{figure}[ht]
\includegraphics[width=\textwidth]{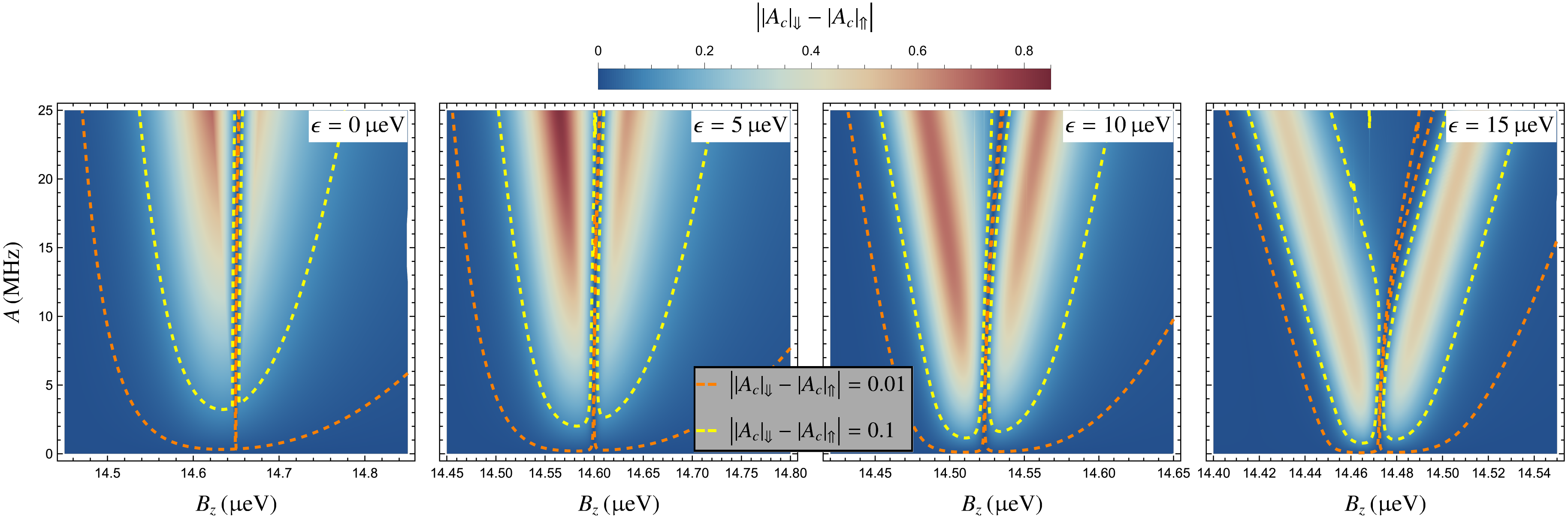}
\caption{Absolute value of the readout contrast as a function of the hyperfine interaction strength $A$ and the homogeneous magnetic field $B_z$ for four values of the detuning $\epsilon$. The remaining parameters are as in Fig. \ref{fig:cavityTransmission}.
For the QD-donor system studied in this paper, we have $A=25\,\mathrm{MHz}$.}
\label{fig:readoutcontrasthyperfine}
\end{figure}

\color{black}
\vspace{4mm}
\section{Effective hyperfine Hamiltonian \label{appendix:EffectiveHnuclearspin}}
The Schrieffer Wolff transformation \eqref{eq:Heffformulanu} results in the Hamiltonian
\begin{align}
 H_{n}=&\frac{E_{\Bar{\nu}}}{2}\nu_z+\left(\Bar{\omega}_c+\frac{\delta E_{\Bar{\nu}}}{2}\nu_z\right)a^{\dagger}a+c_4 \nu_x 
 +(c_5+c_6 \nu_z)\left(a+a^{\dagger}\right)
 +c_{7}\left(\nu^{+}a+\nu^{-}a^{\dagger}\right)
 +g_{\nu}\left(\nu^{+}a^{\dagger}+\nu^{-}a\right)\nonumber\\
 &+\left(c_9+c_{10}\nu_z\right)\left(a^2+a^{\dagger^2}\right) 
 +\left(c_{11}+c_{12}\nu_z\right)\left(a^3+a^{\dagger^3}\right)
+\left(c_{13}+c_{14}\nu_z\right)\left(a^{\dagger}aa+a^{\dagger}a^{\dagger}a\right).
\label{eq:Heff}
\end{align}
The discussion in Sec. \ref{sec:nuclear_spin-photon_coupling} of the main text does not involve a further investigation of the terms proportional to the coefficients $c_i$. Therefore we only give the explicit expressions of $E_{\Bar{\nu}},\,\Bar{\omega}_c,\,\delta E_{\Bar{\nu}}$ and $g_{\nu}$ in terms of the system parameters. 
The diagonal part of \eqref{eq:Heff} is characterized by
\begin{align}
  E_{\Bar{\nu}}=&-\frac{A}{4}(1+\sin\theta)+\frac{\tfrac{1}{8}Ab_x^2\sin^2\theta\,(1+\sin\theta)}{B_z^2-\left(\tfrac{A}{4}(1+\sin\theta)\right)^2}+\frac{\tfrac{1}{8}Ab_x^2\cos^2\theta}{(B_z+\Omega)^2-\left(\tfrac{A}{4}\right)^2}+\frac{\tfrac{1}{2}Ag_c^2\cos^2\theta\,\sin\theta}{(\Omega+\omega_c)^2-\left(\tfrac{A}{4}\sin\theta\right)^2}\nonumber\\
  &+\frac{\tfrac{1}{128}A^3\cos^2\theta\,\sin\theta}{\Omega^2-\left(\tfrac{A}{4}\sin\theta\right)^2}-\frac{\tfrac{1}{16}A^2(1+\sin^2\theta)}{B_z}-\frac{\tfrac{1}{16}A^2\cos^2\theta}{\tfrac{A}{4}\sin\theta-B_z-\Omega}+\mathcal{O}(V^3),\\[10pt]
    \Bar{\omega}_c=&\omega_c+g_c^2\cos^2\theta\,\left(\frac{\omega_c-\Omega}{(\Omega-\omega_c)^2-\left(\tfrac{A}{4}\sin\theta\right)^2}-\frac{\omega_c+\Omega}{(\Omega+\omega_c)^2-\left(\tfrac{A}{4}\sin\theta\right)^2}\right)\nonumber\\
    &+\frac{128 A^2 g_c^2 \cos^2\theta\,\sin^2\theta\,\Omega(16 \Omega^2-A^2\sin^2\theta)}{(A^2\sin^2\theta-16\Omega^2)((A \sin\theta)^2-(4 (\Omega - \omega _c))^2)  ((A \sin\theta)^2-(4 (\Omega +\omega_c))^2)},\\[10pt]
	\delta E_{\Bar{\nu}}=&\frac{16 A g_c^2 \sin\theta\, \cos^2\theta\, \left(A^2 \sin
		^2\theta +48 \Omega ^2-16 \omega_c^2\right)}{((A \sin\theta
		)^2-(4 (\Omega - \omega _c))^2)  ((A \sin \theta )^2-(4 (\Omega +\omega_c))^2)},
	\end{align}
	where the third order contributions in $V$ to $E_{\Bar{\nu}}$ are neglected for simplicity.  
For the nuclear spin-photon coupling $g_{\nu}$ we find:
	\begin{align}
	g_{\nu}=&\tfrac{1}{4} A b_x g_c \cos ^2\theta\, \left[\sin \theta \, \left(\frac{4}{(A \sin \theta +A-4 B_z) (A-4 (B_z+\Omega +\text{$\omega
   $c}))}+\frac{1}{(B_z-\omega_c) (A \sin \theta +4 (B_z+\Omega ))}\right.\right.\nonumber\\
   &\left.\left.+\frac{1}{B_z (A \sin \theta +4 (B_z+\Omega -\text{$\omega
   $c}))}-\frac{8}{(A \sin\theta+4 (B_z+\Omega )) (A \sin \theta+4 (B_z+\Omega -\omega_c))}\right.\right.\nonumber\\
   &\left.\left.+\frac{1}{(B_z-\omega_c) (A
   \sin\theta+4 \Omega -4 \omega_c)}-\frac{4}{(A \sin \theta+4 \Omega -4 \omega_c) (A \sin\theta+4 (B_z+\Omega -\text{$\omega
   $c}))}\right.\right.\nonumber\\
   &\left.\left.+\frac{4}{(A-4 (B_z+\Omega )) (A \sin \theta+A-4 (B_z+\omega_c))}-\frac{4}{(A-4 (B_z+\Omega +\omega_c)) (A \sin
   \theta-4 (\Omega +\omega_c))}\right.\right.\nonumber\\
   &\left.\left.+\frac{4}{(A \sin\theta+A-4 (B_z+\omega_c)) (A \sin\theta-4 (\Omega +\text{$\omega
   $c}))}+\frac{4}{(A-4 \Omega ) (A \sin\theta +A-4 B_z)}\right.\right.\nonumber\\
   &\left.\left.-\frac{4}{(A+4 \Omega ) (A \sin\theta+4 (B_z+\Omega ))}-\frac{8}{(A-4
   (B_z+\Omega )) (A-4 (B_z+\Omega +\omega_c))}\right.\right.\nonumber\\
   &\left.\left.-\frac{4}{(A-4 \Omega ) (A-4 (B_z+\Omega ))}+\frac{1}{A B_z+4 B_z \Omega
   }\right)+\frac{1}{B_z (A \sin \theta +4 (B_z+\Omega -\omega_c))}\right.\nonumber\\
   &\left.+\frac{4}{(A \sin \theta+4 \Omega -4 \omega_c) (A \sin
   \theta +4 (B_z+\Omega -\omega_c))}+\frac{4}{(A-4 (B_z+\Omega )) (A \sin \theta +A-4 (B_z+\omega_c))}\right.\nonumber\\
   &\left.+\frac{4}{(A \sin
   \theta+A-4 (B_z+\omega_c)) (A \sin \theta -4 (\Omega +\omega_c))}+\frac{1}{A B_z+4 B_z \Omega }+\frac{4}{(A-4 \Omega )
   (A-4 (B_z+\Omega ))}\right].
	\end{align}

\end{widetext}

\bibliography{bibliography}

\end{document}